\documentclass[english]{article}	
\usepackage[utf8]{inputenc}				
\usepackage[big,online]{dgruyter}	
\usepackage{lmodern} 
\usepackage{microtype}
\usepackage[numbers,square,sort&compress]{natbib}
\usepackage{color}
\usepackage{ragged2e}
\usepackage{soul}

\DeclareUnicodeCharacter{2009}{\,} 
\usepackage{booktabs}

\theoremstyle{dgthm}

\theoremstyle{dgdef}

\usepackage{multirow}

\begin{document}

	\articletype{Review Article} 
	\received{Month	DD, YYYY}
	\revised{Month	DD, YYYY}
  \accepted{Month	DD, YYYY}
  \journalname{De~Gruyter~Journal}
  \journalyear{YYYY}
  \journalvolume{XX}
  \journalissue{X}
  \startpage{1}
  \aop
  \DOI{xx.xxxx/sample-YYYY-XXXX}

\title{Nanowire-based Integrated Photonics for Quantum Information and Quantum Sensing}

\flushleft

\runningtitle{Nanowire-based Photonics for Quantum Information and Sensing}

\author*[1]{Jin Chang}
\author[2]{Jun Gao}
\author[3]{Iman Esmaeil Zadeh} 
\author[4]{Ali W. Elshaari} 
\author[5]{Val Zwiller} 
\runningauthor{J.~Chang et al.}

\affil[1]{\protect\raggedright Kavli Institute of Nanoscience, Department of Quantum Nanoscience, Delft University of Technology, 2628CJ Delft, the Netherlands. E-mail: j.chang-1@tudelft.nl}
\affil[2]{\protect\raggedright 
Department of Applied Physics, Royal Institute of Technology, Albanova University Centre, Roslagstullsbacken 21, 106 91 Stockholm, Sweden}
\affil[3]{\protect\raggedright 
Department of Imaging Physics (ImPhys), Faculty of Applied Sciences, Delft University of Technology, Delft 2628 CJ, The Netherlands}
\affil[4]{\protect\raggedright 
Department of Applied Physics, Royal Institute of Technology, Albanova University Centre, Roslagstullsbacken 21, 106 91 Stockholm, Sweden}
\affil[5]{\protect\raggedright 
Department of Applied Physics, Royal Institute of Technology, Albanova University Centre, Roslagstullsbacken 21, 106 91 Stockholm, Sweden}


\justify 

\abstract{At the core of quantum photonic information processing and sensing, two major building pillars are single-photon emitters and single-photon detectors. In this review, we systematically summarize the working theory, material platform, fabrication process, and game-changing applications enabled by state-of-the-art quantum dots in nanowire emitters and superconducting nanowire single-photon detectors. Such nanowire-based quantum hardware offers promising properties for modern quantum optics experiments. We highlight several burgeoning quantum photonics applications using nanowires and discuss development trends of integrated quantum photonics. Also, we propose quantum information processing and sensing experiments for the quantum optics community, and future interdisciplinary applications.}

\keywords{photonics integrated  circuits, nanowires, epitaxial quantum dots, superconducting nanowire single photon detector, quantum information processing, quantum sensing.}

\maketitle

\section{Introduction} \label{intro}

Having the highest possible speed allowed by the laws of physics, photons are the fastest carrier to transmit information. The large bandwidth and bosonic behavior, allowing for photons to share (part of) a channel without a short circuit, have made photons the main choice for communication networks. Photonics is also a vital element in the toolbox of future quantum technologies. Conventionally, quantum optics experiments were carried out using tabletop equipment\cite{book_quantum_optics_Aspect,book_quantum_optics_Knight,book_quantum_optics_Fox}. While such setups are flexible, accommodating components based on innovative technologies and material platforms, and have enabled the demonstration of a very impressive prototype to manipulate numerous photons\cite{jiuzhang1}, ultimately this cannot be scaled much further. Over the past 25 years, integrated photonics was established as a reliable, scalable, and cost-efficient alternative to bulk optics \cite{Tanzilli_QIP_review_2011,Flamini_QIP_review_2018,Elshaari_2020_review, Moody_QIP_roadmap_2022} and offers exciting prospects for intense upscaling. \\
Quantum photonics has taken a similar progress path as the broader field of optics, i.e. starting with bulk free-space equipment and miniaturization via integration. Integrated quantum photonics (IQP) has already gone a long way and milestone theoretical and experimental works on IQP have already been reported in different platforms \cite{mile_Carolan_2015,mile_Bartlett_2020,mile_Lodahl_2017,mile_Metcalf2014,mile_Zhang_2014,mile_Wang2020}. However, many challenges still prevent scaling such platforms. As we argue in this paper, the most formidable challenge ahead for IQP is compatibility issues: performances of individual quantum integrated photonics components, i.e. emitters, photonic circuits, and detectors, have excelled in the past two decades \cite{singe_photon_source_and_detectors_review_Eisaman2011,singe_photon_source_review_Meyer2020_multiplexing,snspd_review1}, but these elements are either incompatible or their integration comes at the cost of significant performance penalties. Hybrid integration is a method in which individual elements are created in their compatible platform and environment, and are then transferred to a host substrate. These approaches have gained significant attention, and several independent works have demonstrated the viability and potential of these techniques.

Nanowire-based integrated photonics is an important member of the hybrid integration class and the focus of this paper. We organize this review in the following structure: After a general introduction to integrated quantum photonics in section \ref{intro}, we discuss nanowire-based emitters in section \ref{emitters}, integrated nanowire detectors in section \ref{integrated_detectors} and then review the material and fabrication methodologies in section \ref{materials_and_fab}. Section \ref{apps} summarizes promising integrated quantum photonics applications enabled by nanowire technology and section \ref{outlooks} is dedicated to prospects and promising future applications of nanowire-based, quantum-enhanced photonic technology followed by a conclusion in section \ref{conclusion}.

\section{Quantum emitters} \label{emitters}

Numerous quantum information technologies rely heavily on non-classical light. In particular, several quantum-secured communication and quantum computing techniques need light sources that can generate single photons, entangled photon pairs, or cluster states as a necessary resource. For the creation of non-classical light, sources based on solid-state nanoscale emitters have emerged as a high-quality and potentially scalable option in the last few years\cite{aharonovich2016solid}. A variety of solid-state emitters are utilized to generate non-classical light, including carbon nanotubes\cite{he2018carbon}, semiconductor quantum dots (QDs) \cite{arakawa2020progress}, and 2D materials\cite{toth2019single}. Due to their high emission rate\cite{somaschi2016near}, narrow emission line width\cite{tomm2021bright}, record low multi-photon emission probability\cite{schweickert2018demand}, and high indistinguishability\cite{ding2016demand}, QDs are often used in advanced applications for the demonstration of quantum advantage \cite{wang2017high}. In as-grown planar quantum dot samples, the light extraction efficiency is very limited due to the large refractive index contrast of the host material. Recently, there has been a rapid development in producing single photon sources with superior light extraction efficiency\cite{reimer2012bright}, and small fine structure splitting\cite{singh2009nanowire,versteegh2014observation}, through embedding QDs in a photonic nanowire. In these nanowires, the size and placement of the QDs are well controlled, resulting in excellent spectral purity\cite{laferriere2022approaching}. For an in-depth review of nanowire-based sources of non-classical light sources, we refer the reader to reference \cite{dalacu2019nanowire}. \\

There are two main bottom-up methods to fabricate nanowire-based QDs\cite{hobbs2012semiconductor}: Selective-area epitaxy, where the nanowire is grown on a patterned substrate, and vapor–solid–liquid epitaxy, where the metal catalyst is used to grow the nanowire. Such techniques result in ultra-bright and clean emission from nanowire quantum dots approaching the Fourier-transform limit\cite{laferriere2022approaching} with a Gaussian mode emission profile and near unity coupling to the guided optical mode, enabling high collection efficiency\cite{laferriere2022unity} and large operation bandwidth through controlling the growth conditions. As an example, the typical emission wavelength of InAsP quantum dots in InP nanowires is around 900 nm, by controlling the size of the quantum dot in the growth phase, which sets the confinement potential, the emission wavelength can be extended to the telecom range \cite{haffouz2018bright} while maintaining a Gaussian emission profile for efficient optical fiber coupling\cite{bulgarini2014nanowire_gaussian}.  \\

The uniqueness of the bottom-up growth method of nanowires mitigates a number of possible processes that may cause linewidth broadening. For example, each device can contain a single or several quantum dots in a highly controlled process \cite{laferriere2020multiplexed}. Additionally, dot nucleation can happen without the development of a wetting layer, and the sidewalls of the photonic nanowire are epitaxially formed crystal planes rather than etched using dry or wet etching methods. Although multi-emitter circuits have been realized \cite{elshaari2017chip}, one standing challenge is to tune all the quantum sources to the same operating wavelength for achieving high two-photon interference visibility. Piezoelectric and thermal tuning have been realized experimentally to control the QDs emission wavelength\cite{elshaari2018strain,elshaari2017chip}, and an electrostatic approach through Stark-shift has been recently theoretically proposed\cite{zeeshan2019proposed}. It is still an open question whether only wavelength tuning is sufficient to reach high photon indistinguishability, without the need for Purcell enhancement. On short time scales, the indistinguishability of photons is governed by $T_{2}/2(T_{1})$, where $T_{1}$ is the emitter lifetime and $T_{2}$ is the coherence time. Also, 
$1/T_{2}=1/2T_{1} +T^{*}_{2}$, where $T^{*}_{2}$ is defined as the pure dephasing time deviating from the natural line width \cite{liu2018high}. Recently, by using a highly optimized growth process, the line widths of nanowires QDs emission spectra were shown to be only 2 times of the Fourier transform limit for above-band excitation \cite{laferriere2022approaching}. Although further measurements are needed to characterize indistinguishability through two-photon interference visibility, this important result further demonstrates the great potential of nanowire quantum dots for quantum information processing.  \\

In addition to the attractive single photon emission properties, high-fidelity spin-qubit initialization using optical pumping was recently realized in nanowire quantum dots. Such solid-state qubits can be potentially used for quantum repeaters to generate entanglement between flying and anchored qubits \cite{lagoudakis2016initialization}. Additionally, nanowire quantum dots can produce polarization-entangled photon pairs through the biexciton-exciton cascade \cite{singh2009nanowire,versteegh2014observation}. The process makes use of the Pauli exclusion principle in the quantum dot's s-shell. A completely filled s-shell leads to a zero-spin bound biexciton state, and two cascaded photons are then emitted with zero total angular momentum. Since the photon-pair state cannot be factorized (into a product state of each individual photon's polarization state), the polarization state of the photon pair is a maximally entangled Bell state. Time-bin entanglement was also realized with nanowire quantum dots \cite{aumann2022demonstration}, which may be more suitable for long-distance communication in optical fibers. Additionally, the growth of nanowire quantum dots can be tailored to tune the emission wavelength (e.g. by strain engineering) to interface with atomic memories, which is of paramount importance for quantum memories and quantum repeaters\cite{anderson2022delaying,al2022single,bharadwaj2021interfacing} \cite{Chen_2016_QD_piezzo}.  \\

\begin{figure}[hthp]
\centering
\fbox{\includegraphics[width=0.9\columnwidth]{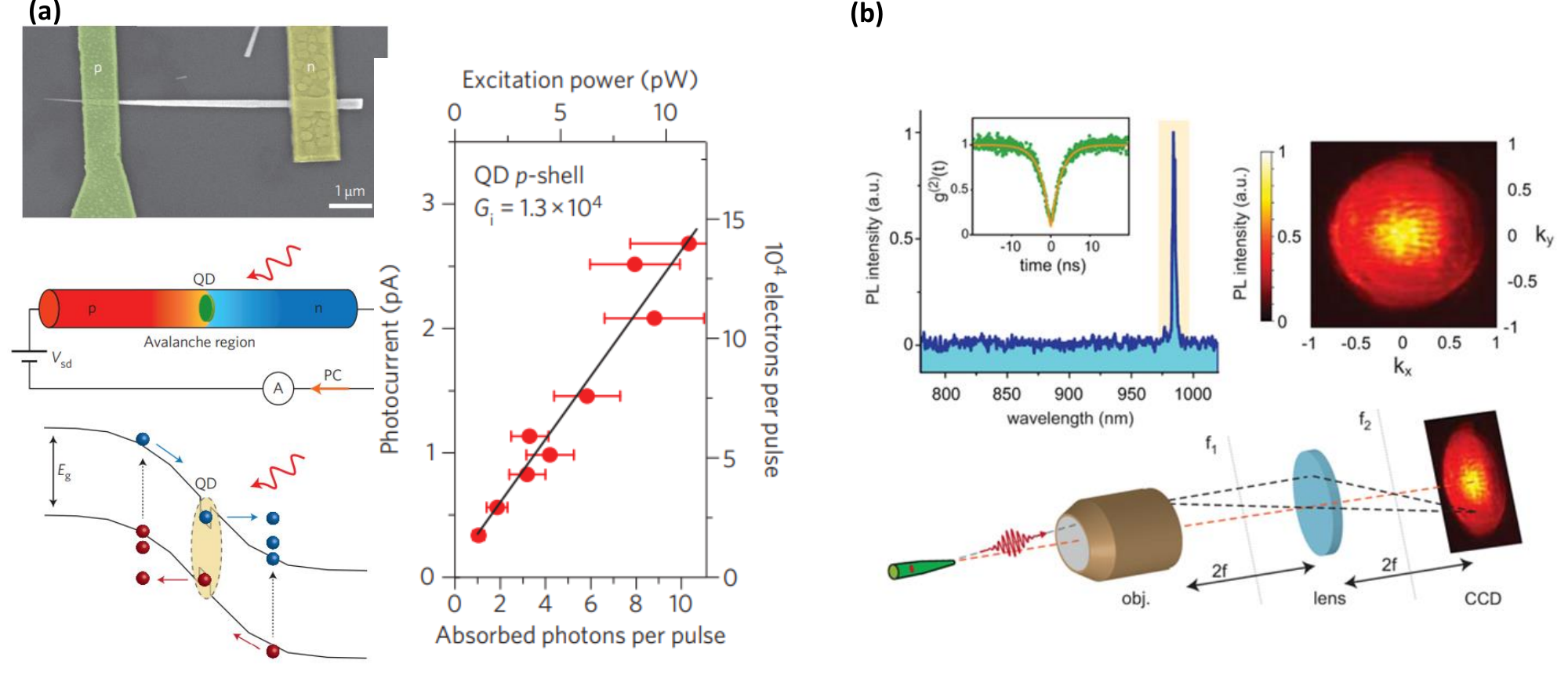}}
\caption{\label{nanowire-QD-transport-and-optics} Nanowires, due to their unique geometry, offer the possibility to both detect and emit light. (a) An embedded QD in a nanowire is contacted and used as an efficient light detector \cite{Gabriele_nw_avalanche_detector}. (b) Nanowires with engineered geometry can efficiently guide light emitted by QDs and beam-shape the emission into a near-perfect Gaussian, compatible with standard single-mode optical fibers \cite{bulgarini2014nanowire_gaussian}.}
\label{fig1_semiconductor_nanowire_detector_and_gaussian_emission}
\end{figure}

Besides promising emission properties, QDs can also be used for single-photon detection \cite{Blakesley:2005_QD_detector}. In principle, QD-based detectors, when designed appropriately, can be sensitive to a wide range of wavelengths, extending to the mid-infrared \cite{Grotevent:2021_QD_graphene}. Recently, semiconductor nanowire-based light detectors have shown very promising detection performances \cite{Gibson2019_NW_detector_reimer}. Nanowires, when utilized as in-plane light detectors (integrated with the plane of photonic circuits), have the added advantage of being compatible with integrated photonics \cite{Cao:2010_nanowire_detector, Gabriele_nw_avalanche_detector}. An example of such a detector is shown in figure \ref{fig1_semiconductor_nanowire_detector_and_gaussian_emission}(a). Semiconductor nanowire detectors still have a long way to go to match the performance metrics of other on-chip single-photon detection technologies (such as superconducting nanowires, which will be covered in section \ref{integrated_detectors}), but thanks to their potential room temperature operation condition and compatibility with monolithic integration techniques, they hold great promises for future integrated quantum photonics.\\

\section{Integrated nanowire single-photon detectors} \label{integrated_detectors}

The inception of superconducting nanowires for single-photon detection dates back to 2001 \cite{gol2001picosecond}. Since then, the superconducting nanowire single-photon detectors (SNSPDs) field has witnessed great progress and improvement, mostly with standard fiber-coupled devices \cite{snspd_review1}. The working principle of on-chip integrated SNSPDs is the same as conventional fiber-coupled detectors \cite{natarajan2012superconducting}, typically superconducting nanowires are DC-current biased below their critical current and temperature T$_c$ and when a photon is absorbed, cooper pairs are broken thus quasi-particles (and/or vortices) are created. This leads to the formation of a normal-conducting region in the wire, redirecting the current toward the readout electronics. After a certain recovery time, the superconducting nanowires return to their superconducting state, and the dynamics of this process depend on the kinetic inductance of the device and the readout circuitry. \\

However, fiber-coupled SNSPDs are not favorable for scaling up detector numbers and lowering the cost per detection channel. For many quantum optics experiments, detectors are usually separately placed in a closed-cycle cryostat, which increases the total cost of targeted applications \cite{ustcbs2017,jiuzhang1,jiuzhang2}. Integrating multiple SNSPDs on-chip would enable large-scale on-chip quantum optics experiments with a more compact chip-scale design and fabrication. \\

\begin{figure}[hthp]
\centering
\fbox{\includegraphics[width=0.9\columnwidth]{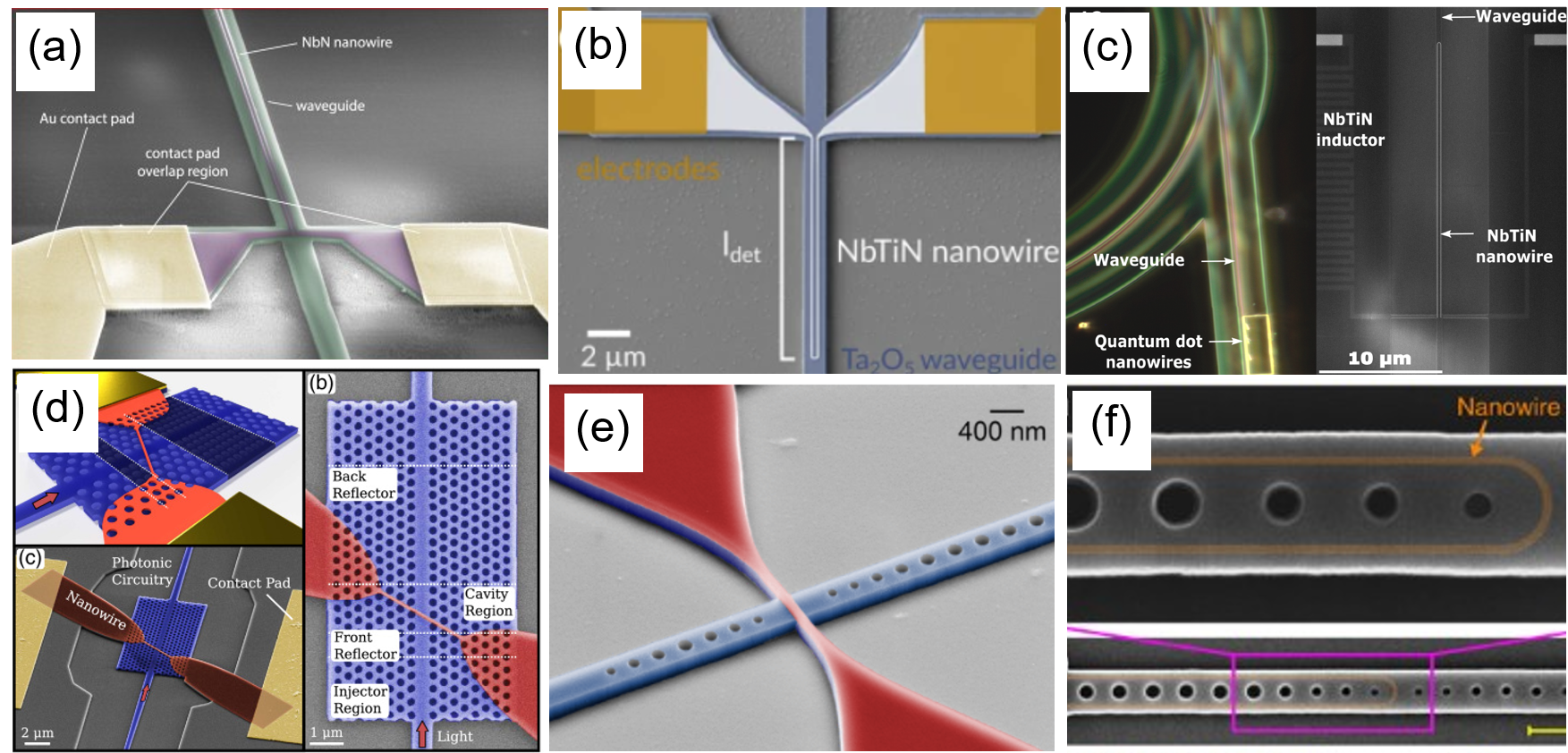}}
\caption{\label{onchip-detectors}Superconducting nanowire on-chip integration with different optical structures: (a) SNSPDs embedded in Silicon Nitride nanophotonic circuits with internal quantum efficiencies close to unity at 1550 nm wavelength \cite{kahl2015waveguide}; (b) NbTiN superconducting nanowire integrated with Ta$_2$O$_5$ waveguide achieving 75\% on-chip detection efficiency at 1550 nm \cite{wolff2020superconducting}; (c) Optical microscope picture of a part of the integrated photonic circuit including quantum dots nanowires, a waveguide, and a ring resonator \cite{gourgues2019controlled}; (d) short NbN superconducting nanowire integrated into two-dimensional double heterostructure photonic crystal cavity with recovery times of 480 ps \cite{2D_cavity}; (e) Nanobeam cavity-integrated NbN superconducting nanowire single-photon detectors with sub-nanosecond decay and recovery times \cite{phc_snspd}, and (f) U-shaped NbTiN nanowires atop silicon-on-insulator waveguides are embedded in asymmetric nanobeam cavities with a near unity on-chip quantum efficiency for 1545 nm \cite{akhlaghi2015waveguide}.}
\end{figure}

In recent years, great efforts have been put into the integration of large numbers of single-photon detectors on-chip. Unlike traditional fiber-coupled SNSPDs \cite{99point5}, in order to achieve high detection efficiency, SNSPDs are either integrated into photonic waveguides with traveling wave geometry \cite{Sprengers2011, pernice2012}, or placed in planar photonic crystal cavities \cite{phc_snspd, 2D_cavity}. SNSPDs are typically placed atop optical waveguides and the traveling light field is absorbed by sufficient long superconducting nanowires. This requires depositing superconducting thin films on top of the waveguide layer and then performing electron beam lithography followed by etching steps; An alternative approach is to embed the superconducting nanowires into optical waveguides. As shown in \cite{gourgues2019controlled}, SNSPDs can first be fabricated and tested, and then the optical waveguides can be deterministically formed (deposited and patterned) to integrate with pre-selected SNSPDs. \\

When designing on-chip SNSPDs, the main concerns include choosing materials, and fabrication routine as described in the following section \ref{materials_and_fab}. Due to the small footprint and lower kinetic inductance of the on-chip SNSPDs, these detectors can exhibit faster (sub-nanosecond \cite{2D_cavity}) recovery time and a lower dark count rate compared to fiber-coupled devices. Also, since the total length of integrated nanowires is significantly shorter, the probability of introducing imperfections is reduced thus a higher yield can be expected. In the future, we expect large numbers of SNSPDs or SNSPD cameras to be integrated into quantum photonic chips for achieving more sophisticated tasks. For a more detailed review of the on-chip integration of SNSPDs with different types of waveguides and their performance, interested researchers are referred to \cite{ferrari2018waveguide}. In the future, besides deploying large numbers of detectors on-chip, how to read out the detection signals of thousands of detection channels remains as an outstanding challenge. In section \ref{QIP}, we propose a hybrid integration architecture, which utilizes both the advantage of photonics and electronics technology to overcome the signal read-out challenge.



\section{Material and fabrication for quantum emitters and detectors} 
\label{materials_and_fab}
In this section, we summarize representative material candidates and the general fabrication process for integrated QDs emitters and superconducting nanowire single-photon detectors. Also, we highlighted different hybrid nanowire device integration approaches, and focus on the integration methods and their challenges and advantages in terms of selectivity and scalability. \\

\subsection{Epitaxial growth of nanowire quantum dot}
Semiconductor quantum dots have been systematically studied as single-photon sources in various quantum optics applications. They are typically fabricated with top-down or bottom-up approaches. Epitaxial methods, for example, molecular-beam epitaxy (MBE) or Metalorganic vapor-phase epitaxy (MOVPE) \cite{joyce1985molecular, leys1981study} are frequently used for growing nanowire QDs. During the epitaxial process, short segments of smaller-band-gap semiconductors are embedded in a larger-band-gap semiconductor. Taking the InAsP QD in InP nanowire as an example \cite{bright_QDs}: InP nanowires were first grown on InP substrate in an MOVPE reactor with Au particles as catalyst and trimethyl-indium plus phosphine as precursors. Afterward, an InAsP quantum dot was incorporated by introducing As in the reactor using an Arsine flux. Afterward, the chamber temperature was raised to favor radial versus axial growth, thus forming the InP shells. By controlling growth time and temperature, the nanowire geometry is shaped with an optimum nanowire diameter and tapering angle towards the tip. With the tapered waveguide structure, such QDs have high photon extraction efficiency. Each QDs nanowire can be individually tested and transferred following optical measurements to select the best quantum dots on photonic chips for more advanced quantum optics measurements and integration in complex architectures. For more detailed material and fabrication methods regarding nanowire QDs, we refer to \cite{mantynen2019single}.  \\

\subsection{Superconducting film deposition and detector integration}
For superconducting nanowire single-photon detectors, the most commonly used film deposition technology is magnetron sputtering. It is a physical vapor deposition (PVD) process, where a magnetically confined plasma is created near the surface of a target material (e.g. Titanium or Niobium). Positively charged energetic ions from the plasma collide with the negatively charged target material, and atoms from the target are “sputtered”, and then deposit onto the substrate \cite{zichi2019optimizing}. A single target made of alloys or multiples targets each containing an elementary material can be used for thin superconducting film deposition. Afterward, with one-step electron beam lithography followed by reactive ion etching, the nanowire pattern is created on different substrates previously chosen for sputtering. An alternate approach is to fabricate and pre-test the superconducting nanowires, then deposit waveguide materials on top and selectively etch them to cover the detectors with waveguides \cite{gourgues2019controlled}. Due to the shorter lengths of integrated superconducting nanowires and thus fewer imperfections, the yield, and recovery time of on-chip detectors can be significantly improved. \\

\subsection{Hybrid integration of quantum emitters and detectors}
Given the variety of the needed building blocks for single photon generation to manipulation and detection, a monolithic material platform will not be sufficient to realize complex photonic systems. Recently, there has been a rapid development in hybrid photonic integration approaches.  Nanowire quantum emitters pioneered the field of hybrid quantum photonic integration\cite{elshaari2020hybrid}. The geometry of nanowires not only allows for efficient light extraction\cite{bright_QDs, bulgarini2014nanowire_gaussian}, single-photon detection \cite{Gabriele_nw_avalanche_detector} and electroluminescence \cite{cavalli_nanowire_el_2016} \ref{nanowire-QD-transport-and-optics}, but also enables their transfer from the growth chip to  other materials and platforms, such as piezoelectric crystals and silicon-based photonic circuits\cite{elshaari2018strain,zadeh2016deterministic,elshaari2017chip,gourgues2019controlled,mnaymneh2020chip,chen2016controlling}. Two methods of transfer can be used: non-selective through dispersing the nanowires randomly on a target sample and then building the photo-electronic circuitry that incorporates the nanowires using lithographic techniques \cite{yang2020proximitized}, or selective site-controlled technique using pick and place transfer\cite{elshaari2018strain}. In the latter, a typical setup  consists of a tungsten needle with a 100 nm tip diameter mounted on a high-precision XYZ stage. The needle can be controlled to adhere to a specific nanowire on the growth chip through van der Waals forces. Then, the tip is used to break the nanowire from the growth chip and transfer it to the target chip with a marker field for further alignment in subsequent fabrication steps. In addition to the pick-and-place technique, transfer printing approaches were also developed \cite{osada2019strongly,katsumi2018transfer}. Using a high-precision positioning system, a rubber stamp composed of polydimethylsiloxane (PDMS) can be used to transfer suspended structures from a growth chip to a target. While pick-and-place and transfer printing techniques offer high selectivity of the target quantum emitter in terms of emission line width and wavelength, the scalability is limited, as each quantum emitter has to be mechanically transferred to the target chip. Another promising approach, which over larger scalability, at the expense of selectivity, is the wafer bonding approach. III-V epitaxially produced QD sources have been successfully bonded to silicon nitride photonic chips using this technology\cite{davanco2017heterogeneous}. Then using mechanical grinding, chemical mechanical polishing, or chemical etching, the sacrificial layer is removed once the bonding has occurred to reveal the photonic circuit layer. For more details about hybrid integration, we refer the readers to \cite{kim2020hybrid,Elshaari_2020_review}. Finally, hybrid integration was not only limited to quantum emitters, recently, SNSPD hybrid integration was also realized with 100$\%$ yield. In the future, interfaces between SNSPDs and external electronic will be necessary for the coupling of SNSPDs with intricate, dynamically reconfigurable photonic structures for active feedback operations.  \\

\begin{table} [htbp]
\centering

\caption{Overview of the nanowire fabrication and materials choice for representative quantum emitters and detectors.}
\begin{tabular}{llll}

\multicolumn{4}{c}{\textbf{Quantum dots in nanowire single-photon emitters}}\\ \cmidrule{2-3}

Materials & Fabrication method & Key performance & Reference\\ 
\midrule

InAs/GaAs  & Solid-source MBE & Emission at 1300 nm & \cite{alloing2005growth}\\ 
InAsP/InP  & Chemical beam epitaxy & Emission 880-1550 nm & \cite{haffouz2018bright} \\
InAsP/InP  & MOVPE & Light-extraction efficiency of 42\% & \cite{bright_QDs}\\
InAsP/InP  & Chemical beam epitaxy & High-fidelity entangled photon-pairs  & \cite{versteegh2014observation}\\
InAsP/InP  & Vapor–liquid-solid (VLS) epitaxy & Multi-photon event <1\%  & \cite{dalacu2012ultraclean}\\
AlGaAs/GaAs  & MBE & Background-free & \cite{schweickert2018demand}\\
GaAsP/GaP  & Low-pressure MOVPE & Bright QDs grown on Si substrate & \cite{borgstrom2005optically}\\
InGaN/GaN  & Plasma-assisted MBE & Electrically driven QDs & \cite{deshpande2013electrically}\\
InGaAs/GaAs& MOCVD & Strain-engineered telecom wavelength DQs & \cite{mrowinski2019excitonic}\\
AlGaN/GaN& MOCVD & Room-temperature operation with $g^{(2)}_0=0.13$ & \cite{holmes2014room}\\
\midrule 

\multicolumn{4}{c}{\textbf{Superconducting nanowire single-photon detectons}} \\ \cmidrule{2-3} 
\textbf{Materials} & \textbf{Fabrication method} & \textbf{Key performance} & \textbf{Reference}\\ 
\midrule
NbN  & Molecular-beam epitaxy (MBE) & Working on AlN ${\chi}^{(2)}$ circuits & \cite{cheng2020epitaxial}\\
NbN  & Atomic layer deposited (ALD) & Working till 2006 nm & \cite{taylor2021infrared}\\
NbTiN  & Magnetron co-sputtering & >99\% efficiency & \cite{99point5}\\
NbTiN  & Magnetron co-sputtering & 7.7ps timing jitter & \cite{esmaeil2020efficient}\\
WSi  & Magnetron co-sputtering &93\% efficiency &\cite{marsili2013detecting} 	\\
MoSi  & Magnetron co-sputtering &>98\% efficiency & \cite{reddy2020superconducting} 	\\
NbRe  & Magnetron sputtering &Visible-infrared detection & \cite{NbRe} 	\\
TaN  & Magnetron sputtering &Large-area X-ray detection & \cite{yang2021large}	\\ 
NbN  & Magnetron sputtering & >98\% efficiency & \cite{hu2020detecting}\\     
MgB$_2$  & Hybrid physical chemical vapor deposition & 130 ps relaxation time & \cite{cherednichenko2021low}\\ 
\end{tabular}
\label{tab:Table1}
\end{table}

\begin{figure}[hthp]
\centering
\fbox{\includegraphics[width=0.7\columnwidth]{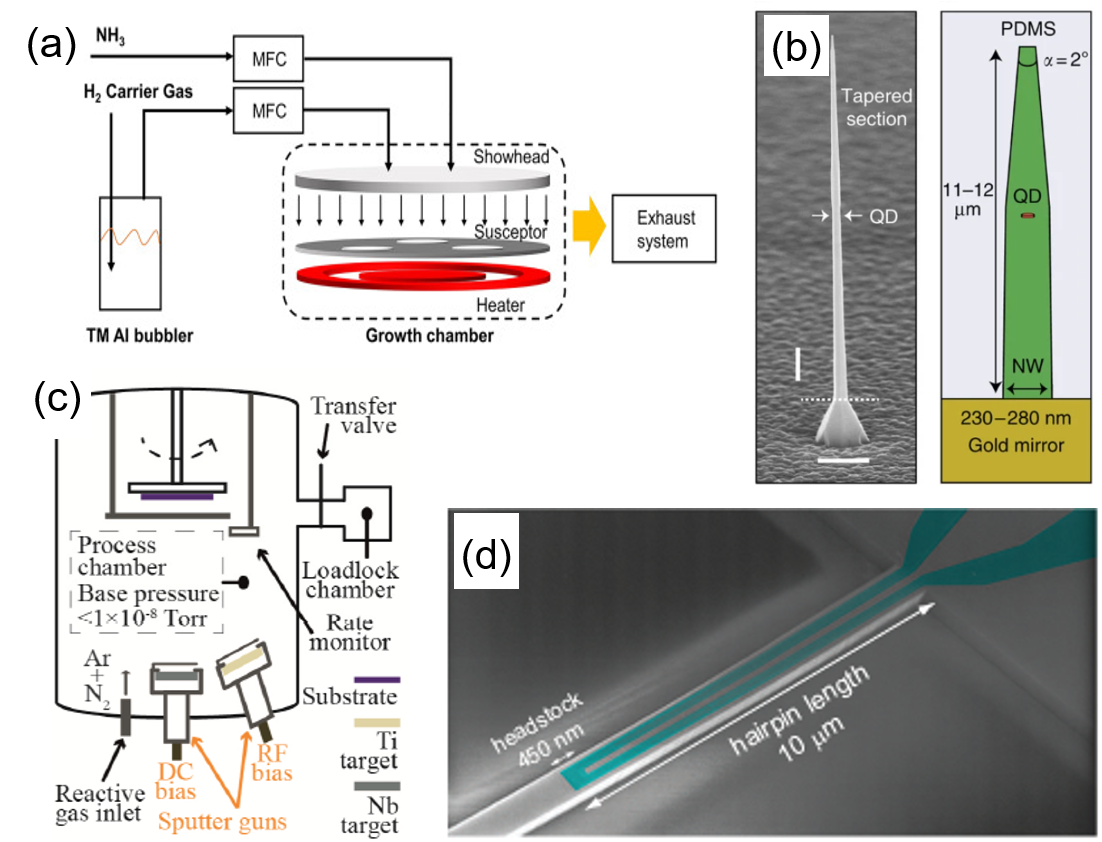}}
\caption{\label{materials-fab} Representative fabrication technologies and device images for quantum emitters and quantum detectors: (a) The schematic of an MOVPE system for growing QDs in nanowire single-photon emitters \cite{marks1982metalorganic}; (b) Left is an SEM image of an InAsP quantum dot embedded in tapered InP nanowire waveguide; Right image is tailored nanowire geometry embedded in polymer with bottom gold mirror \cite{bright_QDs}; (c) Schematics of a magnetron co-sputtering system for depositing superconducting thin films \cite{zichi2019optimizing}, and (d) False-color SEM image of a MoSi hairpin SNSPD on SOI waveguide \cite{li2016nano}.}
\end{figure}


\section{Emerging quantum photonic technology and outlooks} \label{apps}

In this section, we highlight a non-exhaustive number of established and emerging quantum optics applications enabled by (partially) integrating quantum emitters, waveguides, and detectors on-chip. We also present perspectives on future quantum technologies with their benchmarks and targets using integrated quantum photonic technology. 

\subsection{Photonic boson sampling}\label{boson} 

\begin{figure}[htpb]
\centering
\fbox{\includegraphics[width=0.95\columnwidth]{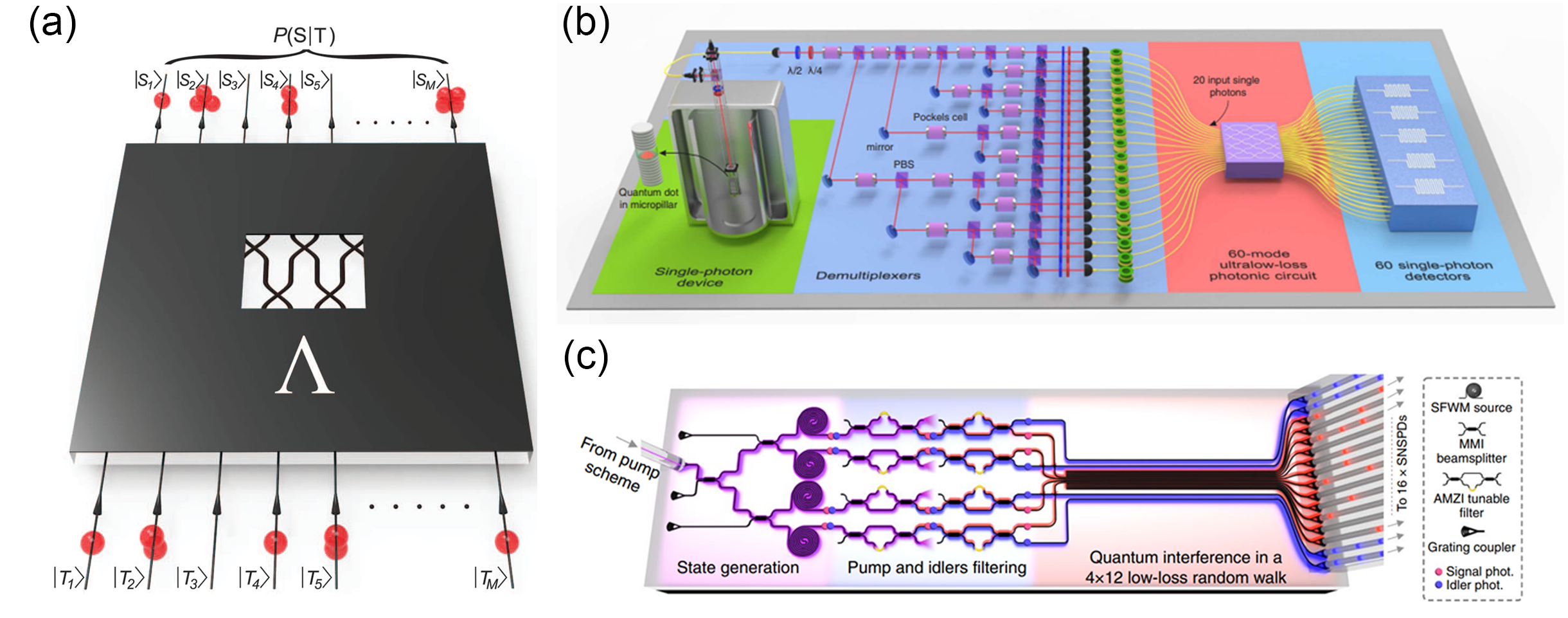}}
\caption{\label{bosonsampling-image} Boson sampling experiments using Fock states and Gaussian states as the input: (a) Schematic of a boson sampling machine: identical bosons are prepared and interfere in a linear optical network, with the output distribution efficiently sampled from the linear network \cite{oxford2013bs}. (b) Indistinguishable photons are generated by a semiconductor quantum dot and de-multiplexed in different spatial modes. The photons interfere in an ultra-low loss bulk crystal and are detected by nanowire single-photon detectors \cite{wang2019boson}. (c) On-chip generation of non-classical Gaussian light. The silicon photonic chip also integrates filters and a passive linear network to perform Gaussian boson sampling \cite{paesani2019generation}.}
\end{figure}

Boson sampling, first proposed by Scott Aaronson and Alex Arkhipov \cite{AA2013}, is a computational task aiming to demonstrate quantum advantage with an intermediate-scale quantum device. The central idea of the task is to sample the output distribution of indistinguishable bosons interfering in a linear network, as schematically shown in figure \ref{bosonsampling-image}(a). With the increase of photon numbers, the task becomes intractable using the classical computation approach due to the intrinsic hardness of calculating the matrix permanent, and thus is considered an excellent candidate to demonstrate quantum computational advantage \cite{bsreview}. To this end, the photonic system is one of the most suitable platforms, as the key elements (quantum emitters, linear state evolution, and single-photon detectors) are widely available with the current technology as described in the previous sections \ref{emitters} and \ref{integrated_detectors}. 

As the early demonstrations of boson sampling, several research groups have chosen to use photon pairs generated from spontaneous parametric down-conversion process and silica photonic chips \cite{oxford2013bs,queensland2013bs,tillmann2013bs,crespi2013bs}. To overcome the low generation rate issue from probabilistic photon sources, some research groups chose semiconductor quantum dot emitter as the input states \cite{ustcbs2017,loredo2017boson}, thus the photon number was dramatically increased from the initial 3 to 20 photons with a state space dimension up to 10$^{14}$ \cite{wang2019boson}. The experimental setup shown in figure \ref{bosonsampling-image}(b) represents the largest scale of boson sampling using quantum dot emitter and SNSPD detection. In the meanwhile, a variant of boson sampling was proposed to use a single mode squeezed vacuum state as the input state instead of the Fock state, called Gaussian boson sampling \cite{hamilton2017gaussian,kruse2019detailed}. Unlike the original  boson sampling protocol, where photon numbers are conserved, Gaussian boson sampling offers a boost in the photon number since the source could emit random numbers of photon pairs. Gaussian boson sampling has been experimentally demonstrated using both ultra-low loss bulk optics \cite{jiuzhang1,jiuzhang2} and silicon photonics platform \cite{paesani2019generation}. Figure \ref{bosonsampling-image}(c) demonstrates the design of the integrated photonics circuit, unlike the bulk optics setup, the chip has a rather small footprint and could be extended to a large-scale device. Nonlinear effects in silicon can naturally generate photon pairs via spontaneous four-wave mixing in either spiral waveguides \cite{paesani2020near} or ring resonators \cite{silverstone2015qubit,vaidya2020broadband}. It is still an open question whether the current technology is capable of scaling boson sampling to arbitrarily large dimensions while maintaining the quantum advantage. Also, such ultra-large-scale experiments require huge numbers of single-photon detectors and the corresponding coincidence detection systems, e.g. distributed SNSPDs with cryogenics and control electronics, making them far from cost-effective \cite{nanophotonics_you}. As a result, integrated photonics technology is commonly believed to be a promising approach to reaching scalable boson sampling \cite{wang2020integrated}. In addition, the optical modes of Gaussian boson sampling can be mapped to vibrational normal modes to solve the vibronic spectrum of a molecule \cite{huh2015boson}. Gaussian boson sampling also holds the potential to solve graph theory problems \cite{bradler2018gaussian,arrazola2018using} and molecular docking for pharmaceutical drug design \cite{banchi2020molecular}. The on-chip photonic circuits potentially possess higher stability (e.g. less phase drift and frequency drift), better isolation from the environment, and low power consumption to be programmable \cite{arrazola2021quantum} to solve the challenging applications we mentioned above. With the advances in integrated photonics technology, nowadays larger modular linear optical circuits \cite{modular-interferometer}, bright quantum emitters \cite{bright_QDs} and controlled integration of detectors \cite{gourgues2019controlled} are more widely available, thus a scalable integrated boson sampler for specific practical problems is the next step to be achieved. \\

\subsection{Quantum walks}\label{walks} 

\begin{figure}[ht]
\centering
\fbox{\includegraphics[width=0.8\columnwidth]{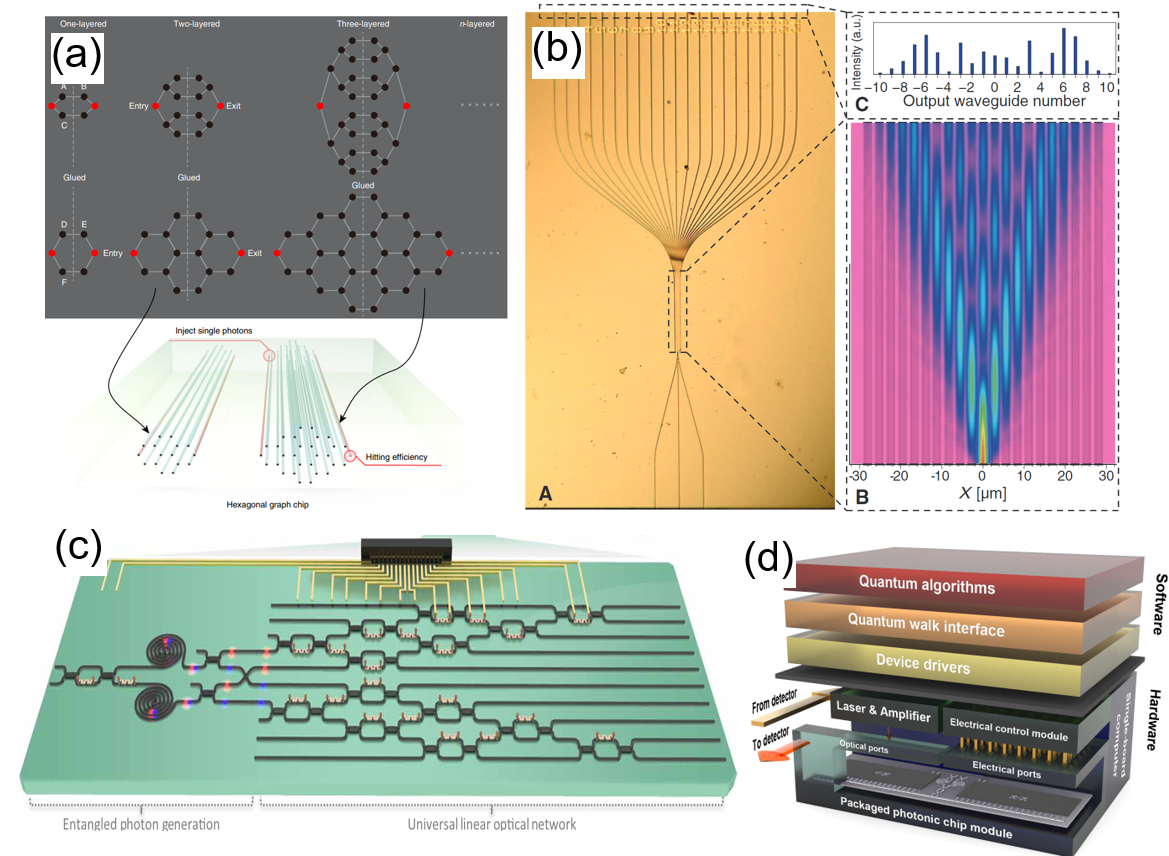}}
\caption{\label{quantum_walk} Representative quantum walk works (a) Top: theoretical graphs of quantum fast-hitting using two-dimensional hexagonal structure; Bottom: implementation of a quantum walk with a glued binary tree structure on photonic chips using femtosecond laser-written waveguide arrays \cite{tang2018fs}; (b) Left: A continuously coupled waveguide array for realizing correlated photon quantum walks with 21-waveguide array; Right: simulation and experiment output pattern of 810-nm laser light propagating through the waveguide array \cite{peruzzo2010quantum}; (c) Graph-theoretic quantum algorithms on a silicon photonic quantum walk processor, including generating spatial-entangled photons and implementing universal five-dimensional unitary process \cite{quantum_walksQiang}, and (d) Schematic of a large-scale full-programmable quantum walk system stack, where software stack compiles quantum algorithms into quantum walk settings and then operated by the hardware \cite{quantum_walksWang}.}
\label{quantum-walks} 
\end{figure}

Quantum walks are the quantum counterparts of classical random walks first proposed in 1993\cite{aharonov1993quantum}, where quantum superposition plays an extremely important role. Unlike a classical particle, a quantum particle can simultaneously propagate in different directions, and this unique behavior leads to the ballistic transport feature of the quantum walk. Normally, there are two different models of quantum walks, namely discrete quantum walks, and continuous quantum walks. Depending on the tasks, quantum walks could provide either exponential (quantum fast-hitting) \cite{childs2003exponential,tang2018fs} or polynomial (quantum search algorithm) \cite{childs2004spatial,benedetti2021quantum} speedup over classical algorithms \cite{quantum_walks_review}, and could even implement universal quantum computation \cite{childs2009universal,childs2013universal}. The photon, which inherently exhibits the wave-particle duality, is naturally a good candidate as a "walker". The quantum properties of photons including superposition, quantum interference, and entanglement can be employed to perform various quantum walk experiments. Early-stage quantum walk experiments are conducted by the bulk optics beam splitters \cite{broome2010discrete} and fiber loops \cite{schreiber2010photons,schreiber20122d} in the time domain. As a comparison, integrated circuits offer higher phase stability, thus further leading to a higher level of integration \cite{peruzzo2010quantum,tang2018experimental,jiao2021two}. For example, figure \ref{quantum-walks}(a) illustrates the implementation of a quantum walk with a glued binary tree structure on photonic chips using femtosecond laser-written waveguide arrays, and figure \ref{quantum-walks}(b) demonstrates a continuously coupled waveguide array for realizing correlated photon quantum walks.  Similarly, in figure \ref{quantum-walks}(c), an integrated photonic platform consisting of reconfigurable linear optical networks and controllable on-chip entangled photon pair sources, demonstrates the simulation of thousands of continuous-time quantum walk evolutions \cite{quantum_walksQiang}. Most recently, a full-stack quantum walk processor based on an integrated photonic chip is used to demonstrate a series of quantum applications, from graph-theoretic applications to quantum simulations of topological phases \cite{quantum_walksWang} as shown in figure \ref{quantum-walks}(d).

\subsection{On-chip quantum communication}\label{QKD}

\begin{figure}[hthp]
\centering
\fbox{\includegraphics[width=0.9\columnwidth]{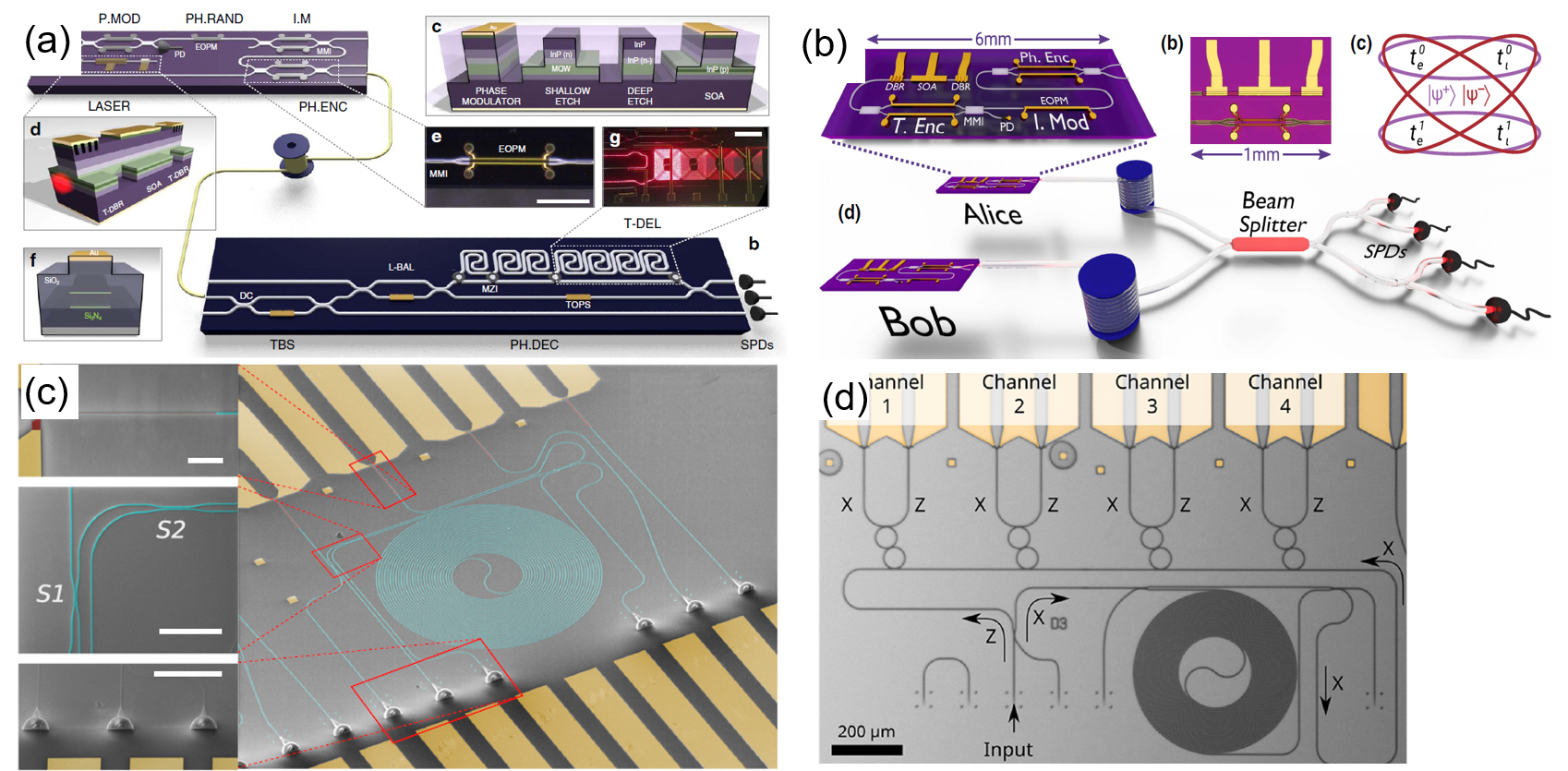}}
\caption{\label{QKD-image} Different integrated QKD systems: (a) Integrated photonic devices for multiprotocol QKD using InP transmitter, continuously tunable laser diode, photodiode, and multimode interferometers \cite{sibson2017chip}; (b) Integrated MDI-QKD using on-chip DBR laser, MZI elements, and nanowire single-photon detectors \cite{semenenko2020chip}; (c) SEM image of QKD receiver chip with optical coupler, waveguide, splitter, optical delay lines, and nanowire single-photon detectors \cite{beutel2021detector}, and (d) SEM image of a four-channel time-bin QKD receiver chip \cite{QKD_optica}.}
\end{figure}

Unconditional secure information exchange is a demanding requirement for both governments and individuals. Over the past few decades, quantum key distribution (QKD) with security fundamentally guaranteed by the laws of physics, has grown rapidly from lab demonstrations to the deployment of commercially available systems connecting distant cities. Demonstrated by the recently launched quantum satellite, space-to-ground QKD has linked locations over 1200 km apart \cite{yin2017satellite,liao2017satellite,ren2017ground}, and even intercontinental quantum communication over 7600 km \cite{liao2018satellite} has been  realized. In general, a QKD system includes a signal-sending part for generating required photon states and a signal-receiving part for photon state detection. Integration efforts have been made to lower the size, cost, and energy consumption of both ends. Ideally, a "sender chip" using integrated light sources with polarization modulators, phase modulators, and power attenuators could efficiently generate certain photon states (e.g. BB84 with polarization or phase encoding), while a "receiver chip" integrated with many single-photon detectors, optical circuits, and control electronics could register the signal photons to decode the information \cite{paraiso2021photonic,semenenko2020chip}. There have been several experimental demonstrations of silicon photonic transmitters for polarization \cite{ma2016silicon}, time-bins \cite{sibson2017integrated,sibson2017chip} and space division multiplexing encoding \cite{ding2017high}. An intercity metropolitan QKD test was performed using a silicon photonics encoder, reaching a quantum communication distance over 42 km \cite{bunandar2018metropolitan}. Another recent experiment using silicon photonics realized the chip-based transmitter and receiver for continuous-variable QKD \cite{zhang2019integrated}. In a recent demonstration, a four-channel silicon nitride-based integrated QKD receiver achieved a total secret-key rate of up to 12.17 Mbit/s at a 3.35 GHz clock rate using wavelength-division de-multiplexing and waveguide-integrated superconducting nanowire single-photon detectors \cite{QKD_optica}. Besides QKD, chip-based quantum teleportation has also been demonstrated using an integrated photonics platform \cite{llewellyn2020chip}. For more detailed QKD-related protocols, implementation, security analysis, and attacking risks, the reader can refer to \cite{QKD_RMP1,QKD_RMP2}.

\subsection{Optical neural networks for machine learning}\label{computing}

In the era of big data, artificial intelligence has greatly revolutionized the modern world and has applications in many areas, for example, image and language analysis, self-driving vehicles, and the famous alpha Go \cite{silver2017mastering}. Currently, electronic circuits are still the prevailing computing power support for artificial intelligence, especially promoted by GPU calculations, however, the Von Neumann architecture cannot meet the increasing demand for ultra-large-scale information processing, limited by energy consumption and electronic interference \cite{sengupta2014power,schwabe1998electronic}. Light, as an excellent information carrier, which travels with fast speed and high parallelism, can solve electronic defects and the research of optical neural network (ONN) can boost the development of artificial intelligence with energy and time efficiency \cite{wetzstein2020inference,liu2021research}.

\begin{figure}[hthp]
\centering
\fbox{\includegraphics[width=0.95\columnwidth]{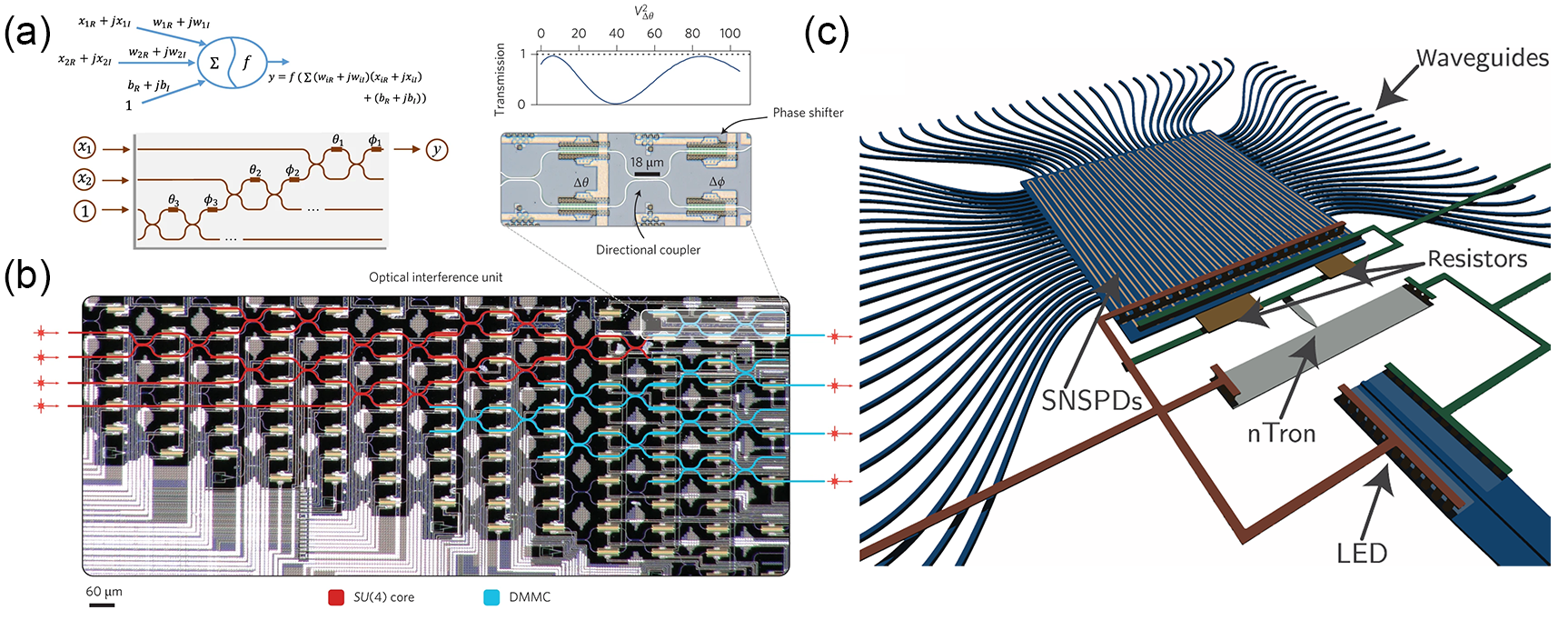}}
\caption{\label{ONN-image} Optical neural networks with integrated photonic devices: (a) Schematic of a complex-valued neuron and its implementation based on integrated MZIs \cite{zhang2021optical}. (b) A micrograph illustration of an optical neural network that can perform both matrix multiplication (red circuit) and attenuation (blue circuit). The enlarged figure is an example of MZI which tunes the internal phase \cite{shen2017deep}. (c) Schematic overview of a stingray neuron using SNSPDs and LED, which is compatible with large-scale networks \cite{shainline2017superconducting}.}
\end{figure}

The optical implementation of neural networks basically contains two parts: linear operation and nonlinear activation, which can be seen as linear multiplication and summation operations, as shown in figure \ref{ONN-image}(a), where a complex-valued neuron is implemented by a mesh of MZI \cite{zhang2021optical}. In a fully connected linear network, each neuron in the output layer is a weighted sum of all input neurons — which can be mathematically represented as a matrix-vector multiplication. Such multiply-accumulate operations can be experimentally implemented by meshes of Mach–Zehnder interferometers (MZIs), as in reference \cite{shen2017deep} and figure \ref{ONN-image}(b). The central idea is to use the principle of interference to implement linear operations, with the tunability of phase shifters in MZIs, the ONN can implement any operation on the input states. Normally the demanding resources (such as MZIs) for a $N$ dimension input is $N^2$, recently, space-efficient integrated diffractive cells are demonstrated \cite{zhu2022space} to further reduce footprint and energy consumption. Nowadays, on-chip ONNs have been extensively realized for the prediction of molecular properties \cite{lau2022photonic}, graph representation learning \cite{yan2022all}, noise-resilient learning \cite{mourgias2022noise}, bacterial foraging training \cite{cong2022chip}, and image classification \cite{ashtiani2022chip}. In recent research, in situ training of ONNs was realized by a fully-integrated coherent optical neural network, including integrated coherent transmitter, matrix multiplication unit, nonlinear function unit and on-chip detection \cite{bandyopadhyay2022single}. To push the physical limit of energy efficiency, a novel spiking neural network was recently demonstrated to perform neuromorphic computing\cite{shainline2017superconducting,buckley2020integrated} using the combination of integrated photonics and SNSPDs, as shown in figure \ref{ONN-image}(c). Such a device has the potential to perform 10 times more operations compared to the human brain with much less energy cost. Another great advance in the field of optical neural networks is the emergence of quantum machine learning \cite{schuld2015introduction,biamonte2017quantum}. With unique quantum features like superposition and entanglement, quantum machine learning algorithms can outperform their classical counterpart with faster speed and fewer resources \cite{haug2021large,huang2021power,krenn2020computer,lamata2020quantum,cerezo2022challenges}.

\subsection{Integrated quantum Lidar system}\label{lidar}

Light detection and ranging, known as "Lidar", is a powerful technology for environmental monitoring, remote target recognition, forest mapping on the earth's surface, and sea fog measurements on ocean\cite{lidar1550,lidar2,lidar3,lidar4}. It detects scattered or reflected light to acquire distance or depth information of remote targets. Typically, a Lidar system consists of pulse laser sources, beam splitters, transceivers, time-correlated single-photon counting electronics, and photodetectors. With the increase in measurement distance, after tens of kilometers, only a few photons can travel back to the detection end, thus the use of single photon detectors can efficiently improve Lidar systems' detection range, depth accuracy, and acquisition time. The superconducting nanowire single-photon detectors developed in the past decades with high efficiency, low timing jitter, and dark count rates \cite{snspd_review1} are becoming the popular choice for recent Lidar systems. A comprehensive review of SNSPD-based Lidars can be found in \cite{lidar-review1}. In the coming future, photon number resolving detectors and efficient mid-infrared detectors will open new detection windows and capabilities for Lidar systems \cite{mid-infrared}. Also, with the development of high-power on-chip laser, photonic integrated circuits, and detector technology, monolithic Lidar chips would enable more compact, space-compatible Lidar applications in the future \cite{lidar5, lidar6}.

\subsection{Meta-surface for integrated quantum optics circuits}\label{meta_ctrl}

With the improvement of high-precision nanofabrication, recent years have seen great progress and increased interest in the field of metasurfaces, which typically contain periodic sub-wavelength metallic/dielectric structures that resonantly couple to the electric and magnetic fields of the light wave \cite{meta1,meta2}. Metasurfaces offer unique solutions to realize unconventional phenomena, for example, negative refraction, achromatic focusing, and electromagnetic cloaking \cite{meta3,meta4,meta5}. The applications of metasurfaces have also been extended from traditional optics to quantum optics, where single photons sources, entangled photons, and single-photon detection are fundamentally required. For the quantum dots emitters described in section \ref{emitters}, the random photon emission issue compromises their use and especially hinders the on-demand manipulation of their spin states. As shown in \cite{meta-QDs}
integrating QDs with metasurface leads to on-demand
generation and separation of the spin states of the emitted single photons along any arbitrary engineered direction. Also, Purcell enhancement can be realized using meta-surface with QDs \cite{vaskin2019light}. When combining metasurface with superconducting nanowire single-photon detectors as described in section \ref{integrated_detectors}, the functionality of these quantum detectors can be greatly extended, for example, achieving spectrum reconstruction on chip \cite{zheng2022photon,xiao2022superconducting}. In the future, combining metasurface with quantum optics elements on-chip will offer new possibilities for controlling single-photon emission, single-photon state manipulation, and single-photon detection and imaging. For more detailed integrated metasurface applications in quantum optics, we refer to \cite{meta-review1}.  

\begin{figure}[hthp]
\centering
\fbox{\includegraphics[width=0.8\columnwidth]{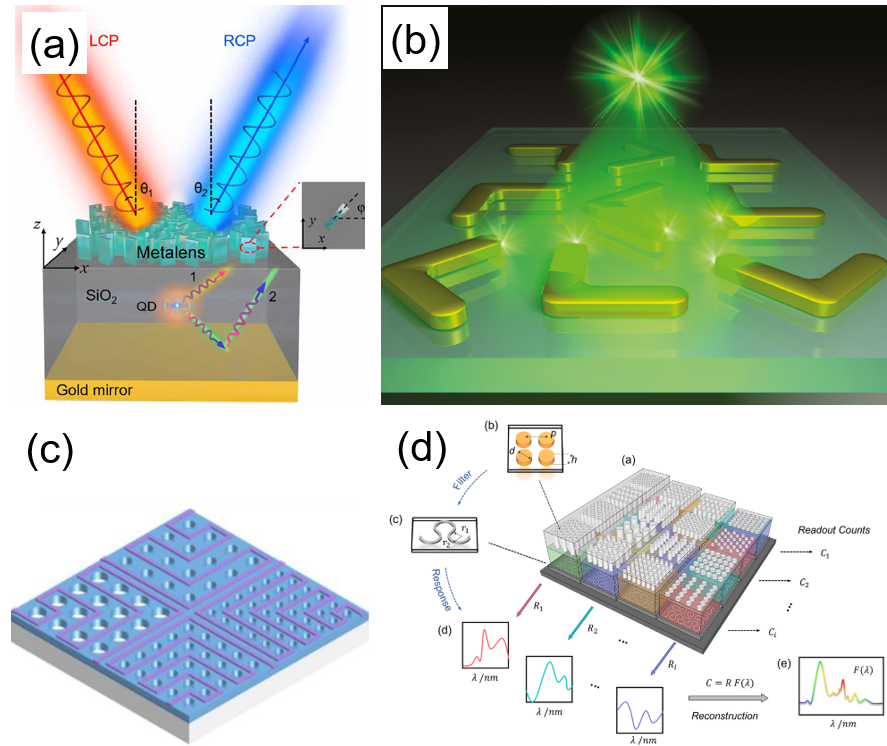}}
\caption{\label{meta} Integration of metasurface with quantum emitters or quantum detectors. (1) Manipulating QD emission using metasurface to achieve on-demand spin state control \cite{meta-QDs}; (b) Illustration of a single quantum emitter interacting with metasurface for Purcell enhancement \cite{vaskin2019light}; (c) Single-photon spectrometer using metasurface array with superconducting nanowire deployed in the region between periodic holes \cite{zheng2022photon}, and (d) computational spectrometer consisting of a 4×4 superconducting nanowire arrays and 3D-printed metasurface \cite{xiao2022superconducting}.}
\end{figure}

\section{Outlooks on future quantum photonics technologies}\label{outlooks}

After decades of development, integrated (quantum) photonics—the science and technology of generating, controlling, and detecting photons on a chip scale-has benefited different industries and society. For example, in telecommunications, where bandwidth and security are greatly demanded and photonic integrated circuits (PICs) offer a viable solution; Other emerging application areas, including quantum photonic computing, bio-photonics sensing, environmental monitoring, and disease diagnosis are also witnessing game-changing breakthroughs triggered by the rapid development of quantum integrated photonic technology. Here, we present two promising envisioned photonic circuit experiments that could further boost the impact of nanowire-based integrated photonics in science and technology.

\subsection{On-chip quantum information processing}\label{QIP}

\begin{figure}[hthp]
\centering
\fbox{\includegraphics[width=\columnwidth]{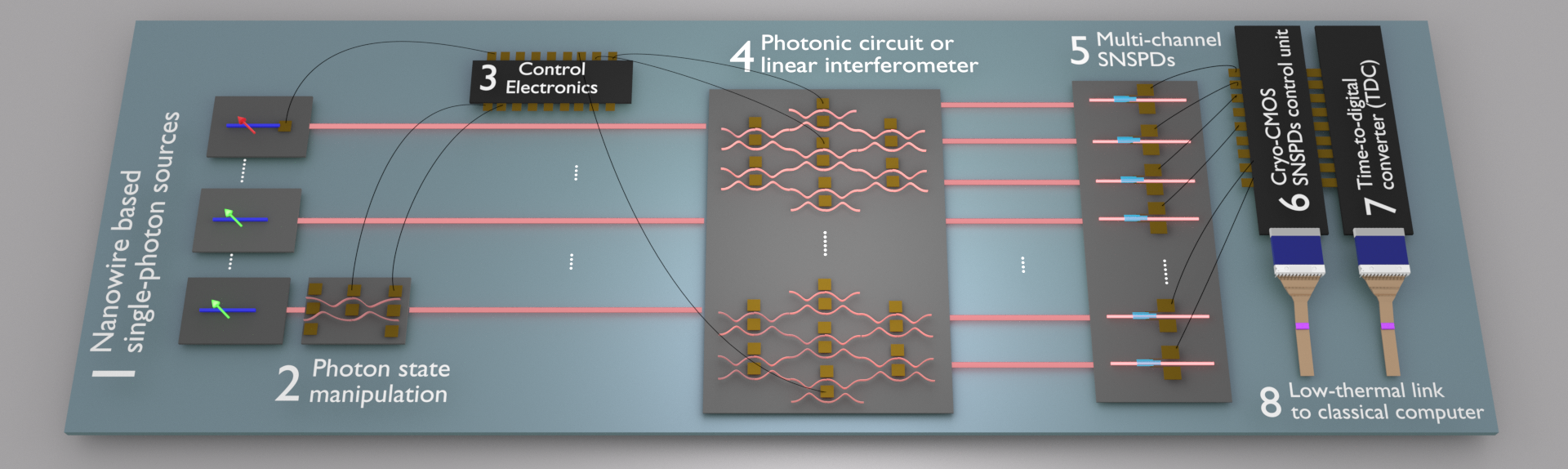}}
\caption{\label{example1}Illustration of an integrated nanowire-based photonic chip to be developed in the future. In this figure, optically excited nanowire QDs sources are shown in green arrows while the electrically pumped single-photon source is shown in red arrows; Optical connections between different components are highlighted in pink lines, and electrical contacts are represented by orange squares; Waveguide-integrated superconducting nanowire single-photon detectors are marked by cyan wires, and integrated control electronic circuits, for example, cryo-CMOS or TDCs, are shown in black rectangles with multiple bonding pads.}
\end{figure}

Recently, the design and production of integrated photonics started to merge into the mainstream of the electronic industry. Such hybrid chips \cite{elshaari2020hybrid} take the complementary advantages of both platforms to perform sophisticated tasks. As illustrated in figure \ref{example1}, the hybrid photonic/electronic integrated circuit contains optically pumped (green) and electrically driven (red) nanowire QDs as single-photon sources. After emission, the photon states can be tuned by additional on-chip elements (e.g. phase shifters \cite{pagani2014tunable}). Multiple photons with precisely controlled initial states are then ejected into the linear interferometer to interact with different photonic quantum computing or simulation protocols. Afterward, the output results (photons) are registered by multi-channels on-chip SNSPDs with integrated control CMOS circuits and time-to-digital (TDC) converters. Such a hybrid chip can be mounted in a compact cryostat without using many coaxial cables, which helps to improve system scalability and reduce the total heat load. In the future, to realize mass production of such proposed hybrid chips, each containing millions of elements, electronics, CMOS-compatible optics, and dedicated superconductor fabrication are simultaneously needed at a foundry level. Also, automatic pre-testing equipment of such chips is significantly important to be developed (e.g. cryogenic probe station \cite{russell2012cryogenic}), where artificial intelligence algorithms can also help to improve failure analysis \cite{pan2021transfer}.       

\subsection{On-chip quantum bio-sensing}\label{Qsensing}

 Different classes of single-photon emitters have been employed for bio-photonic imaging and spectroscopy. Each of these emitters has its own advantages and weaknesses. To this end, quantum dots were among the first emitters to be explored and have already come a long way. Most commonly in biomedical applications and fluorescence microscopy, quantum dots (and in general single-photon emitters) are utilized as biomarkers \cite{Michalet_2005_QD_Review, Chinnathambi:2019_QD_bio_review,Jingjing_2013_QD_bio_review,FARZIN2021_QD_bio_review,Boissiere2013_QD_bio_review} in which their position (localization), color, brightness, lifetime, etc. is linked to a certain biological/chemical factor. Among the important remaining challenges ahead of bio-quantum sensing with quantum dots are chemical toxicity and optical attenuation, i.e. the light from the emitter is highly attenuated by the tissue before reaching the detection optics. Infrared quantum dots can benefit from enhanced transparency of the tissue. Synthesis of high-quality infrared emitting quantum dots \cite{Brunetti2018_infrared_QD_imaging,GIL2021_infrared_QD_for_bio} as well as efficient and precise detection of those photons \cite{Fei_QD_SNSPD_brainimaging_1300_2021,Wang_brain_QD_SNSPD_3rdopticalwindow_2022} have achieved promising results but require further progress. High quantum yield emitters as well as sensitive and low noise detectors with a large active area can further boost the impact of infrared bio-imaging. \\
 
 As described in the previous sections, integrated photonic circuits can generate, transmit, and detect a broad band of optical signals on-chip. This naturally offers the ability to simultaneously detect and identify different biological objects (e.g. single-virus \cite{ozcelik2015optofluidic}, proteins \cite{pelton2000spectroscopic}, single-molecule \cite{pieczonka2008single}), which is one of the key requirements for disease diagnostic and drug developments.\\ 

\begin{figure}[htbp]
\centering
\fbox{\includegraphics[width=\columnwidth]{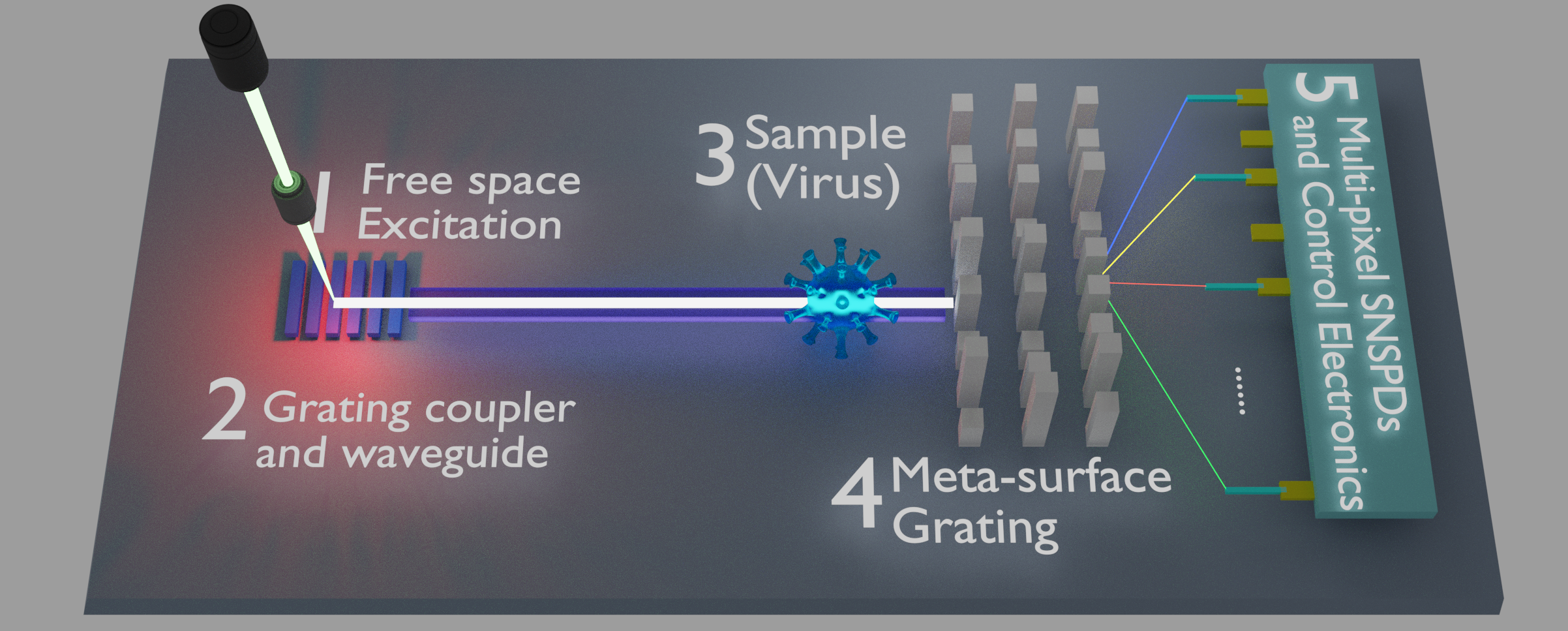}}
\caption{\label{example2}A proposed integrated quantum sensing chip for single-virus testing, including free space excitation laser, grating couplers, optical waveguide, the sample under test (single-virus as an example, can also be a single-molecule or protein, etc.), meta-surface grating, and multi-pixel SNSPDs.}
\end{figure}

Integrated photonic circuits can help to create such a highly sensitive, multi-functional platform on a chip scale. Also, by using single-photon detectors, the sensitivity of such systems can be improved to reach their quantum limits. As depicted in figure \ref{example2}, a broadband excitation light signal transmits from free space to the chip with the help of efficient grating couplers. The waveguide delivers light to the samples under test  (a virus in this case). Afterward, the transmitted or scattered photon signal pass through the meta-surface grating to acquire spectral information with multi-pixel SNSPDs at the detection ports. With precise spectral information, the identity or structure of bio-samples can be efficiently acquired. Such chips hold great potential in both scientific labs and biochemical industries, and again, nanowire-based devices are the key enabling elements in the proposed systems.     

\section{Conclusion and future perspectives} \label{conclusion}

With the 2022 Nobel physics prize awarded to the scientists who opened the quantum optics field using single-photons and entangled photon pairs, the field of quantum photonics science is expected to gain more attention. After decades of development, this field has witnessed great developments in both fundamental theory and real-world applications. The applications of nanowires QDs and SNSPDs are extending from proof-of-principle demonstrations to large-scale quantum computing, quantum simulation, and quantum sensing. Integrated and hybrid quantum photonic solutions are promising approaches for developing next-generation quantum hardware, where the advantages of photonics, electronics, and condensed matter physics can be combined for game-changing innovations. Looking into the future, there are still important challenges to be addressed:

\begin{itemize}
\item From quantum emitters' perspective, To interface the flying qubits with the current optical fiber network, quantum emitters must be developed that operate at the telecommunication wavelength of 1550 nm. A further benefit of this advancement is enabling hybrid integration to silicon on insulator photonics, the most developed photonic platform for both classical and quantum applications. Additionally, one key objective is to interface several nanowire QDs, so that they may be used as solid-state quantum memories, and each may function as a quantum node for communication purposes. Such target application would require Fourier-limited emission and more stringent control over the emission wavelength spread in nanowire quantum dot samples.
Lastly, higher photon indistinguishability without time-gating or post-processing is key for several applications highlighted in this review.
\end{itemize}

\begin{itemize}
\item On the detection side, significant advancements have recently been made, allowing SNSPDs to be used in new fields including biological imaging. Following this advancement, it may be advantageous to realize SNSPDs at longer wavelength ranges. This should be done in collaboration with researchers developing biological markers to maximize the detectors' qualities in the desired wavelength range. Additionally, even though large-scale SNSPD arrays have made significant progress, additional work is still required to realize SNDPD 2D cameras with individual pixel-readout circuitry solutions. In the realms of optical imaging, sensing, and biology, this will be a game-changer. Also, a better understanding of how the superconducting films' material properties influence their photon detection performance will help to improve SNSPDs' yield, and ultimately bring down the cost of this technology. This will advance the commercialization of SNSPDs and expand their use to as-yet-untapped markets and research areas. 
\end{itemize}

To conclude, through a systematic review of the theory, material platform, and fabrication process of the nanowires, this paper serves as a solid reference for both young and senior researchers in the integrated quantum photonics field. Additionally, we propose promising quantum photonics architectures and sensing experiments and aim at attracting a wider range of readers and invoking more collaborations between researchers in different fields.



\bibliographystyle{unsrt}
\bibliography{sample.bib} 

\begin{thebibliography}{100}

\bibitem{book_quantum_optics_Aspect}
Gilbert Grynberg, Alain Aspect, Claude Fabre, and Claude Cohen-Tannoudji.
\newblock {\em Introduction to Quantum Optics: From the Semi-classical Approach
  to Quantized Light}.
\newblock Cambridge University Press, 2010.

\bibitem{book_quantum_optics_Knight}
Christopher Gerry and Peter Knight.
\newblock {\em Introductory Quantum Optics}.
\newblock Cambridge University Press, 2004.

\bibitem{book_quantum_optics_Fox}
M.~Fox.
\newblock {\em Quantum Optics: An Introduction}.
\newblock Oxford Master Series in Physics. OUP Oxford, 2006.

\bibitem{jiuzhang1}
Han-Sen Zhong, Hui Wang, Yu-Hao Deng, Ming-Cheng Chen, Li-Chao Peng, Yi-Han
  Luo, Jian Qin, Dian Wu, Xing Ding, Yi~Hu, et~al.
\newblock Quantum computational advantage using photons.
\newblock {\em Science}, 370(6523):1460--1463, 2020.

\bibitem{Tanzilli_QIP_review_2011}
S.~Tanzilli, A.~Martin, F.~Kaiser, M.P. De~Micheli, O.~Alibart, and D.B.
  Ostrowsky.
\newblock On the genesis and evolution of integrated quantum optics.
\newblock {\em Laser \& Photonics Reviews}, 6(1):115--143, 2012.

\bibitem{Flamini_QIP_review_2018}
Fulvio Flamini, Nicol{\`{o}} Spagnolo, and Fabio Sciarrino.
\newblock Photonic quantum information processing: a review.
\newblock {\em Reports on Progress in Physics}, 82(1):016001, nov 2018.

\bibitem{Elshaari_2020_review}
Ali~W. Elshaari, Wolfram Pernice, Kartik Srinivasan, Oliver Benson, and Val
  Zwiller.
\newblock Hybrid integrated quantum photonic circuits.
\newblock {\em Nature Photonics}, 14(5):285--298, May 2020.

\bibitem{Moody_QIP_roadmap_2022}
Galan Moody, Volker~J Sorger, Daniel~J Blumenthal, Paul~W Juodawlkis, William
  Loh, Cheryl Sorace-Agaskar, Alex~E Jones, Krishna~C Balram, Jonathan C~F
  Matthews, Anthony Laing, Marcelo Davanco, Lin Chang, John~E Bowers, Niels
  Quack, Christophe Galland, Igor Aharonovich, Martin~A Wolff, Carsten Schuck,
  Neil Sinclair, Marko Lon{\v{c}}ar, Tin Komljenovic, David Weld, Shayan
  Mookherjea, Sonia Buckley, Marina Radulaski, Stephan Reitzenstein, Benjamin
  Pingault, Bartholomeus Machielse, Debsuvra Mukhopadhyay, Alexey Akimov,
  Aleksei Zheltikov, Girish~S Agarwal, Kartik Srinivasan, Juanjuan Lu, Hong~X
  Tang, Wentao Jiang, Timothy~P McKenna, Amir~H Safavi-Naeini, Stephan
  Steinhauer, Ali~W Elshaari, Val Zwiller, Paul~S Davids, Nicholas Martinez,
  Michael Gehl, John Chiaverini, Karan~K Mehta, Jacquiline Romero, Navin~B
  Lingaraju, Andrew~M Weiner, Daniel Peace, Robert Cernansky, Mirko Lobino,
  Eleni Diamanti, Luis~Trigo Vidarte, and Ryan~M Camacho.
\newblock 2022 roadmap on integrated quantum photonics.
\newblock {\em Journal of Physics: Photonics}, 4(1):012501, jan 2022.

\bibitem{mile_Carolan_2015}
Jacques Carolan, Christopher Harrold, Chris Sparrow, Enrique Martín-López,
  Nicholas~J. Russell, Joshua~W. Silverstone, Peter~J. Shadbolt, Nobuyuki
  Matsuda, Manabu Oguma, Mikitaka Itoh, Graham~D. Marshall, Mark~G. Thompson,
  Jonathan C.~F. Matthews, Toshikazu Hashimoto, Jeremy~L. O’Brien, and
  Anthony Laing.
\newblock Universal linear optics.
\newblock {\em Science}, 349(6249):711--716, 2015.

\bibitem{mile_Bartlett_2020}
Ben Bartlett and Shanhui Fan.
\newblock Universal programmable photonic architecture for quantum information
  processing.
\newblock {\em Phys. Rev. A}, 101:042319, Apr 2020.

\bibitem{mile_Lodahl_2017}
Peter Lodahl.
\newblock Quantum-dot based photonic quantum networks.
\newblock {\em Quantum Science and Technology}, 3(1):013001, oct 2017.

\bibitem{mile_Metcalf2014}
Benjamin~J. Metcalf, Justin~B. Spring, Peter~C. Humphreys, Nicholas
  Thomas-Peter, Marco Barbieri, W.~Steven Kolthammer, Xian-Min Jin, Nathan~K.
  Langford, Dmytro Kundys, James~C. Gates, Brian~J. Smith, Peter G.~R. Smith,
  and Ian~A. Walmsley.
\newblock Quantum teleportation on a photonic chip.
\newblock {\em Nature Photonics}, 8(10):770--774, Oct 2014.

\bibitem{mile_Zhang_2014}
P.~Zhang, K.~Aungskunsiri, E.~Mart\'{\i}n-L\'opez, J.~Wabnig, M.~Lobino, R.~W.
  Nock, J.~Munns, D.~Bonneau, P.~Jiang, H.~W. Li, A.~Laing, J.~G. Rarity, A.~O.
  Niskanen, M.~G. Thompson, and J.~L. O'Brien.
\newblock Reference-frame-independent quantum-key-distribution server with a
  telecom tether for an on-chip client.
\newblock {\em Phys. Rev. Lett.}, 112:130501, Apr 2014.

\bibitem{mile_Wang2020}
Jianwei Wang, Fabio Sciarrino, Anthony Laing, and Mark~G. Thompson.
\newblock Integrated photonic quantum technologies.
\newblock {\em Nature Photonics}, 14(5):273--284, May 2020.

\bibitem{singe_photon_source_and_detectors_review_Eisaman2011}
M.~D. Eisaman, J.~Fan, A.~Migdall, and S.~V. Polyakov.
\newblock Invited review article: Single-photon sources and detectors.
\newblock {\em Review of Scientific Instruments}, 82(7):071101, 2011.

\bibitem{singe_photon_source_review_Meyer2020_multiplexing}
Evan Meyer-Scott, Christine Silberhorn, and Alan Migdall.
\newblock Single-photon sources: Approaching the ideal through multiplexing.
\newblock {\em Review of Scientific Instruments}, 91(4):041101, 2020.

\bibitem{snspd_review1}
Iman Esmaeil~Zadeh, J~Chang, Johannes~WN Los, Samuel Gyger, Ali~W Elshaari,
  Stephan Steinhauer, Sander~N Dorenbos, and Val Zwiller.
\newblock Superconducting nanowire single-photon detectors: A perspective on
  evolution, state-of-the-art, future developments, and applications.
\newblock {\em Applied Physics Letters}, 118(19):190502, 2021.

\bibitem{aharonovich2016solid}
Igor Aharonovich, Dirk Englund, and Milos Toth.
\newblock Solid-state single-photon emitters.
\newblock {\em Nature Photonics}, 10(10):631--641, 2016.

\bibitem{he2018carbon}
X~He, H~Htoon, SK~Doorn, WHP Pernice, F~Pyatkov, R~Krupke, A~Jeantet,
  Y~Chassagneux, and C~Voisin.
\newblock Carbon nanotubes as emerging quantum-light sources.
\newblock {\em Nature materials}, 17(8):663--670, 2018.

\bibitem{arakawa2020progress}
Yasuhiko Arakawa and Mark~J Holmes.
\newblock Progress in quantum-dot single photon sources for quantum information
  technologies: A broad spectrum overview.
\newblock {\em Applied Physics Reviews}, 7(2):021309, 2020.

\bibitem{toth2019single}
Milos Toth and Igor Aharonovich.
\newblock Single photon sources in atomically thin materials.
\newblock {\em Annual review of physical chemistry}, 70:123--142, 2019.

\bibitem{somaschi2016near}
Niccolo Somaschi, Valerian Giesz, Lorenzo De~Santis, JC~Loredo, Marcelo~P
  Almeida, Gaston Hornecker, S~Luca Portalupi, Thomas Grange, Carlos Anton,
  Justin Demory, et~al.
\newblock Near-optimal single-photon sources in the solid state.
\newblock {\em Nature Photonics}, 10(5):340--345, 2016.

\bibitem{tomm2021bright}
Natasha Tomm, Alisa Javadi, Nadia~Olympia Antoniadis, Daniel Najer,
  Matthias~Christian L{\"o}bl, Alexander~Rolf Korsch, R{\"u}diger Schott,
  Sascha~Ren{\'e} Valentin, Andreas~Dirk Wieck, Arne Ludwig, et~al.
\newblock A bright and fast source of coherent single photons.
\newblock {\em Nature Nanotechnology}, 16(4):399--403, 2021.

\bibitem{schweickert2018demand}
Lucas Schweickert, Klaus~D J{\"o}ns, Katharina~D Zeuner, Saimon~Filipe Covre~da
  Silva, Huiying Huang, Thomas Lettner, Marcus Reindl, Julien Zichi, Rinaldo
  Trotta, Armando Rastelli, et~al.
\newblock On-demand generation of background-free single photons from a
  solid-state source.
\newblock {\em Applied Physics Letters}, 112(9):093106, 2018.

\bibitem{ding2016demand}
Xing Ding, Yu~He, Z-C Duan, Niels Gregersen, M-C Chen, S~Unsleber, Sebastian
  Maier, Christian Schneider, Martin Kamp, Sven H{\"o}fling, et~al.
\newblock On-demand single photons with high extraction efficiency and
  near-unity indistinguishability from a resonantly driven quantum dot in a
  micropillar.
\newblock {\em Physical review letters}, 116(2):020401, 2016.

\bibitem{wang2017high}
Hui Wang, Yu~He, Yu-Huai Li, Zu-En Su, Bo~Li, He-Liang Huang, Xing Ding,
  Ming-Cheng Chen, Chang Liu, Jian Qin, et~al.
\newblock High-efficiency multiphoton boson sampling.
\newblock {\em Nature Photonics}, 11(6):361--365, 2017.

\bibitem{reimer2012bright}
Michael~E Reimer, Gabriele Bulgarini, Nika Akopian, Mo{\"\i}ra Hocevar,
  Maaike~Bouwes Bavinck, Marcel~A Verheijen, Erik~PAM Bakkers, Leo~P
  Kouwenhoven, and Val Zwiller.
\newblock Bright single-photon sources in bottom-up tailored nanowires.
\newblock {\em Nature communications}, 3(1):1--6, 2012.

\bibitem{singh2009nanowire}
Ranber Singh and Gabriel Bester.
\newblock Nanowire quantum dots as an ideal source of entangled photon pairs.
\newblock {\em Physical review letters}, 103(6):063601, 2009.

\bibitem{versteegh2014observation}
Marijn~AM Versteegh, Michael~E Reimer, Klaus~D J{\"o}ns, Dan Dalacu, Philip~J
  Poole, Angelo Gulinatti, Andrea Giudice, and Val Zwiller.
\newblock Observation of strongly entangled photon pairs from a nanowire
  quantum dot.
\newblock {\em Nature communications}, 5(1):1--6, 2014.

\bibitem{laferriere2022approaching}
Patrick Laferri{\`e}re, Aria Yin, Edith Yeung, Leila Kusmic, Marek Korkusinski,
  Payman Rasekh, David~B Northeast, Sofiane Haffouz, Jean Lapointe, Philip~J
  Poole, et~al.
\newblock Approaching transform-limited photons from nanowire quantum dots
  excited above-band.
\newblock {\em arXiv preprint arXiv:2208.00066}, 2022.

\bibitem{dalacu2019nanowire}
Dan Dalacu, Philip~J Poole, and Robin~L Williams.
\newblock Nanowire-based sources of non-classical light.
\newblock {\em Nanotechnology}, 30(23):232001, 2019.

\bibitem{hobbs2012semiconductor}
Richard~G Hobbs, Nikolay Petkov, and Justin~D Holmes.
\newblock Semiconductor nanowire fabrication by bottom-up and top-down
  paradigms.
\newblock {\em Chemistry of Materials}, 24(11):1975--1991, 2012.

\bibitem{laferriere2022unity}
Patrick Laferri{\`e}re, Edith Yeung, Isabelle Miron, David~B Northeast, Sofiane
  Haffouz, Jean Lapointe, Marek Korkusinski, Philip~J Poole, Robin~L Williams,
  and Dan Dalacu.
\newblock Unity yield of deterministically positioned quantum dot single photon
  sources.
\newblock {\em Scientific Reports}, 12(1):1--9, 2022.

\bibitem{haffouz2018bright}
Sofiane Haffouz, Katharina~D Zeuner, Dan Dalacu, Philip~J Poole, Jean Lapointe,
  Daniel Poitras, Khaled Mnaymneh, Xiaohua Wu, Martin Couillard, Marek
  Korkusinski, et~al.
\newblock Bright single inasp quantum dots at telecom wavelengths in
  position-controlled inp nanowires: The role of the photonic waveguide.
\newblock {\em Nano letters}, 18(5):3047--3052, 2018.

\bibitem{bulgarini2014nanowire_gaussian}
Gabriele Bulgarini, Michael~E Reimer, Maaike Bouwes~Bavinck, Klaus~D J{\"o}ns,
  Dan Dalacu, Philip~J Poole, Erik~PAM Bakkers, and Val Zwiller.
\newblock Nanowire waveguides launching single photons in a gaussian mode for
  ideal fiber coupling.
\newblock {\em Nano letters}, 14(7):4102--4106, 2014.

\bibitem{laferriere2020multiplexed}
Patrick Laferriere, Edith Yeung, Lambert Giner, Sofiane Haffouz, Jean Lapointe,
  Geof~C Aers, Philip~J Poole, Robin~L Williams, and Dan Dalacu.
\newblock Multiplexed single-photon source based on multiple quantum dots
  embedded within a single nanowire.
\newblock {\em Nano letters}, 20(5):3688--3693, 2020.

\bibitem{elshaari2017chip}
Ali~W Elshaari, Iman~Esmaeil Zadeh, Andreas Fognini, Michael~E Reimer, Dan
  Dalacu, Philip~J Poole, Val Zwiller, and Klaus~D J{\"o}ns.
\newblock On-chip single photon filtering and multiplexing in hybrid quantum
  photonic circuits.
\newblock {\em Nature communications}, 8(1):1--8, 2017.

\bibitem{elshaari2018strain}
Ali~W Elshaari, Efe B{\"u}y{\"u}k{\"o}zer, Iman~Esmaeil Zadeh, Thomas Lettner,
  Peng Zhao, Eva Sch{\"o}, Samuel Gyger, Michael~E Reimer, Dan Dalacu, Philip~J
  Poole, et~al.
\newblock Strain-tunable quantum integrated photonics.
\newblock {\em Nano letters}, 18(12):7969--7976, 2018.

\bibitem{zeeshan2019proposed}
Mohd Zeeshan, Nachiket Sherlekar, Arash Ahmadi, Robin~L Williams, and Michael~E
  Reimer.
\newblock Proposed scheme to generate bright entangled photon pairs by
  application of a quadrupole field to a single quantum dot.
\newblock {\em Physical review letters}, 122(22):227401, 2019.

\bibitem{liu2018high}
Feng Liu, Alistair~J Brash, John O’Hara, Luis~MPP Martins, Catherine~L
  Phillips, Rikki~J Coles, Benjamin Royall, Edmund Clarke, Christopher Bentham,
  Nikola Prtljaga, et~al.
\newblock High purcell factor generation of indistinguishable on-chip single
  photons.
\newblock {\em Nature nanotechnology}, 13(9):835--840, 2018.

\bibitem{lagoudakis2016initialization}
Konstantinos~G Lagoudakis, Peter~L McMahon, Kevin~A Fischer, Shruti Puri, Kai
  M{\"u}ller, Dan Dalacu, Philip~J Poole, Michael~E Reimer, Val Zwiller,
  Yoshihisa Yamamoto, et~al.
\newblock Initialization of a spin qubit in a site-controlled nanowire quantum
  dot.
\newblock {\em New Journal of Physics}, 18(5):053024, 2016.

\bibitem{aumann2022demonstration}
Philipp Aumann, Maximilian Prilm{\"u}ller, Florian Kappe, Laurin Ostermann, Dan
  Dalacu, Philip~J Poole, Helmut Ritsch, Wolfgang Lechner, and Gregor Weihs.
\newblock Demonstration and modeling of time-bin entangled photons from a
  quantum dot in a nanowire.
\newblock {\em AIP Advances}, 12(5):055115, 2022.

\bibitem{anderson2022delaying}
Paul Anderson, Rubayet Al~Maruf, Sreesh Venuturumilli, Divya Bharadwaj, Sonell
  Malik, Jiawei Qiu, Yujia Yuan, Philip Poole, Dan Dalacu, Michael Reimer,
  et~al.
\newblock Delaying tunable single photons from a quantum dot with an atomic
  ensemble.
\newblock In {\em APS Division of Atomic, Molecular and Optical Physics Meeting
  Abstracts}, volume 2022, pages F01--047, 2022.

\bibitem{al2022single}
Rubayet Al~Maruf, Sai~Sreesh Venuturumilli, Divya Bharadwaj, Paul Anderson,
  Jiawei Qiu, Yujia Yuan, Behrooz Semnani, Sonell Malik, Mohd Zeeshan, Dan
  Dalacu, et~al.
\newblock Single-photon source based on a quantum dot emitting at cesium
  wavelength.
\newblock In {\em Optical and Quantum Sensing and Precision Metrology II},
  volume 12016, pages 239--246. SPIE, 2022.

\bibitem{bharadwaj2021interfacing}
Divya Bharadwaj, Paul Anderson, Sreesh Venuturumilli, Rubayet Al~Maruf, Jiawei
  Qiu, Taehyun Yoon, Behrooz Semnani, Yujia Yuan, Jinjin Du, Mohd Zeeshan,
  et~al.
\newblock Interfacing quantum dots with laser-cooled atomic ensembles.
\newblock In {\em Optical and Quantum Sensing and Precision Metrology}, volume
  11700, pages 211--217. SPIE, 2021.

\bibitem{Chen_2016_QD_piezzo}
Yan Chen, Iman~Esmaeil Zadeh, Klaus~D. Jöns, Andreas Fognini, Michael~E.
  Reimer, Jiaxiang Zhang, Dan Dalacu, Philip~J. Poole, Fei Ding, Val Zwiller,
  and Oliver~G. Schmidt.
\newblock Controlling the exciton energy of a nanowire quantum dot by strain
  fields.
\newblock {\em Applied Physics Letters}, 108(18):182103, 2016.

\bibitem{Blakesley:2005_QD_detector}
J.~C. Blakesley, P.~See, A.~J. Shields, B.~E. Kardyna\l{}, P.~Atkinson,
  I.~Farrer, and D.~A. Ritchie.
\newblock Efficient single photon detection by quantum dot resonant tunneling
  diodes.
\newblock {\em Phys. Rev. Lett.}, 94:067401, Feb 2005.

\bibitem{Grotevent:2021_QD_graphene}
Matthias~J. Grotevent, Claudio~U. Hail, Sergii Yakunin, Dominik Bachmann,
  Michel Calame, Dimos Poulikakos, Maksym~V. Kovalenko, and Ivan Shorubalko.
\newblock Colloidal hgte quantum dot/graphene phototransistor with a spectral
  sensitivity beyond 3 µm.
\newblock {\em Advanced Science}, 8(6):2003360, 2021.

\bibitem{Gibson2019_NW_detector_reimer}
Sandra~J. Gibson, Brad van Kasteren, Burak Tekcan, Yingchao Cui, Dick van Dam,
  Jos E.~M. Haverkort, Erik P. A.~M. Bakkers, and Michael~E. Reimer.
\newblock Tapered inp nanowire arrays for efficient broadband high-speed
  single-photon detection.
\newblock {\em Nature Nanotechnology}, 14(5):473--479, May 2019.

\bibitem{Cao:2010_nanowire_detector}
Linyou Cao, Joon-Shik Park, Pengyu Fan, Bruce Clemens, and Mark~L. Brongersma.
\newblock Resonant germanium nanoantenna photodetectors.
\newblock {\em Nano Letters}, 10(4):1229--1233, 2010.
\newblock PMID: 20230043.

\bibitem{Gabriele_nw_avalanche_detector}
Gabriele Bulgarini, Michael~E Reimer, Mo{\"\i}ra Hocevar, Erik P A~M Bakkers,
  Leo~P Kouwenhoven, and Val Zwiller.
\newblock Avalanche amplification of a single exciton in a semiconductor
  nanowire.
\newblock {\em Nature Photonics}, 6(7):455--458, jul 2012.

\bibitem{gol2001picosecond}
GN~Gol’Tsman, O~Okunev, G~Chulkova, A~Lipatov, A~Semenov, K~Smirnov,
  B~Voronov, A~Dzardanov, C~Williams, and Roman Sobolewski.
\newblock Picosecond superconducting single-photon optical detector.
\newblock {\em Applied physics letters}, 79(6):705--707, 2001.

\bibitem{natarajan2012superconducting}
Chandra~M Natarajan, Michael~G Tanner, and Robert~H Hadfield.
\newblock Superconducting nanowire single-photon detectors: physics and
  applications.
\newblock {\em Superconductor science and technology}, 25(6):063001, 2012.

\bibitem{ustcbs2017}
Hui Wang, Yu~He, Yu-Huai Li, Zu-En Su, Bo~Li, He-Liang Huang, Xing Ding,
  Ming-Cheng Chen, Chang Liu, Jian Qin, et~al.
\newblock High-efficiency multiphoton boson sampling.
\newblock {\em Nature Photonics}, 11(6):361--365, 2017.

\bibitem{jiuzhang2}
Han-Sen Zhong, Yu-Hao Deng, Jian Qin, Hui Wang, Ming-Cheng Chen, Li-Chao Peng,
  Yi-Han Luo, Dian Wu, Si-Qiu Gong, Hao Su, et~al.
\newblock Phase-programmable gaussian boson sampling using stimulated squeezed
  light.
\newblock {\em Physical review letters}, 127(18):180502, 2021.

\bibitem{kahl2015waveguide}
Oliver Kahl, Simone Ferrari, Vadim Kovalyuk, Gregory~N Goltsman, Alexander
  Korneev, and Wolfram~HP Pernice.
\newblock Waveguide integrated superconducting single-photon detectors with
  high internal quantum efficiency at telecom wavelengths.
\newblock {\em Scientific reports}, 5(1):1--11, 2015.

\bibitem{wolff2020superconducting}
Martin~A Wolff, Simon Vogel, Lukas Splitthoff, and Carsten Schuck.
\newblock Superconducting nanowire single-photon detectors integrated with
  tantalum pentoxide waveguides.
\newblock {\em Scientific Reports}, 10(1):1--9, 2020.

\bibitem{gourgues2019controlled}
Ronan Gourgues, Iman~Esmaeil Zadeh, Ali~W Elshaari, Gabriele Bulgarini,
  Johannes~WN Los, Julien Zichi, Dan Dalacu, Philip~J Poole, Sander~N Dorenbos,
  and Val Zwiller.
\newblock Controlled integration of selected detectors and emitters in photonic
  integrated circuits.
\newblock {\em Optics express}, 27(3):3710--3716, 2019.

\bibitem{2D_cavity}
Julian M{\"u}nzberg, Andreas Vetter, Fabian Beutel, Wladick Hartmann, Simone
  Ferrari, Wolfram~HP Pernice, and Carsten Rockstuhl.
\newblock Superconducting nanowire single-photon detector implemented in a 2d
  photonic crystal cavity.
\newblock {\em Optica}, 5(5):658--665, 2018.

\bibitem{phc_snspd}
Andreas Vetter, Simone Ferrari, Patrik Rath, Rasoul Alaee, Oliver Kahl, Vadim
  Kovalyuk, Silvia Diewald, Gregory~N Goltsman, Alexander Korneev, Carsten
  Rockstuhl, et~al.
\newblock Cavity-enhanced and ultrafast superconducting single-photon
  detectors.
\newblock {\em Nano letters}, 16(11):7085--7092, 2016.

\bibitem{akhlaghi2015waveguide}
Mohsen~K Akhlaghi, Ellen Schelew, and Jeff~F Young.
\newblock Waveguide integrated superconducting single-photon detectors
  implemented as near-perfect absorbers of coherent radiation.
\newblock {\em Nature communications}, 6(1):1--8, 2015.

\bibitem{99point5}
J~Chang, JWN Los, JO~Tenorio-Pearl, N~Noordzij, R~Gourgues, A~Guardiani,
  JR~Zichi, SF~Pereira, HP~Urbach, Val Zwiller, et~al.
\newblock Detecting telecom single photons with 99.5- 2.07+ 0.5\% system
  detection efficiency and high time resolution.
\newblock {\em APL Photonics}, 6(3):036114, 2021.

\bibitem{Sprengers2011}
JP~Sprengers, A~Gaggero, D~Sahin, S~Jahanmirinejad, G~Frucci, F~Mattioli,
  R~Leoni, Johannes Beetz, M~Lermer, M~Kamp, et~al.
\newblock Waveguide superconducting single-photon detectors for integrated
  quantum photonic circuits.
\newblock {\em Applied Physics Letters}, 99(18):181110, 2011.

\bibitem{pernice2012}
Wolfram~HP Pernice, C~Schuck, O~Minaeva, M~Li, GN~Goltsman, AV~Sergienko, and
  HX~Tang.
\newblock High-speed and high-efficiency travelling wave single-photon
  detectors embedded in nanophotonic circuits.
\newblock {\em Nature communications}, 3(1):1--10, 2012.

\bibitem{ferrari2018waveguide}
Simone Ferrari, Carsten Schuck, and Wolfram Pernice.
\newblock Waveguide-integrated superconducting nanowire single-photon
  detectors.
\newblock {\em Nanophotonics}, 7(11):1725--1758, 2018.

\bibitem{joyce1985molecular}
BA~Joyce.
\newblock Molecular beam epitaxy.
\newblock {\em Reports on Progress in Physics}, 48(12):1637, 1985.

\bibitem{leys1981study}
MR~Leys and H~Veenvliet.
\newblock A study of the growth mechanism of epitaxial gaas as grown by the
  technique of metal organic vapour phase epitaxy.
\newblock {\em Journal of Crystal Growth}, 55(1):145--153, 1981.

\bibitem{bright_QDs}
Michael~E Reimer, Gabriele Bulgarini, Nika Akopian, Mo{\"\i}ra Hocevar,
  Maaike~Bouwes Bavinck, Marcel~A Verheijen, Erik~PAM Bakkers, Leo~P
  Kouwenhoven, and Val Zwiller.
\newblock Bright single-photon sources in bottom-up tailored nanowires.
\newblock {\em Nature communications}, 3(1):1--6, 2012.

\bibitem{mantynen2019single}
Henrik M{\"a}ntynen, Nicklas Anttu, Zhipei Sun, and Harri Lipsanen.
\newblock Single-photon sources with quantum dots in iii--v nanowires.
\newblock {\em Nanophotonics}, 8(5):747--769, 2019.

\bibitem{zichi2019optimizing}
Julien Zichi, Jin Chang, Stephan Steinhauer, Kristina Von~Fieandt, Johannes~WN
  Los, Gijs Visser, Nima Kalhor, Thomas Lettner, Ali~W Elshaari, Iman~Esmaeil
  Zadeh, et~al.
\newblock Optimizing the stoichiometry of ultrathin nbtin films for
  high-performance superconducting nanowire single-photon detectors.
\newblock {\em Optics express}, 27(19):26579--26587, 2019.

\bibitem{elshaari2020hybrid}
Ali~W Elshaari, Wolfram Pernice, Kartik Srinivasan, Oliver Benson, and Val
  Zwiller.
\newblock Hybrid integrated quantum photonic circuits.
\newblock {\em Nature Photonics}, 14(5):285--298, 2020.

\bibitem{cavalli_nanowire_el_2016}
Alessandro Cavalli, Jia Wang, Iman Esmaeil~Zadeh, Michael~E. Reimer, Marcel~A.
  Verheijen, Martin Soini, Sebastien~R. Plissard, Val Zwiller, Jos E.~M.
  Haverkort, and Erik P. A.~M. Bakkers.
\newblock High-yield growth and characterization of ⟨100⟩ inp p–n diode
  nanowires.
\newblock {\em Nano Letters}, 16(5):3071--3077, 2016.
\newblock PMID: 27045232.

\bibitem{zadeh2016deterministic}
Iman~Esmaeil Zadeh, Ali~W Elshaari, Klaus~D J{\"o}ns, Andreas Fognini, Dan
  Dalacu, Philip~J Poole, Michael~E Reimer, and Val Zwiller.
\newblock Deterministic integration of single photon sources in silicon based
  photonic circuits.
\newblock {\em Nano Letters}, 16(4):2289--2294, 2016.

\bibitem{mnaymneh2020chip}
Khaled Mnaymneh, Dan Dalacu, Joseph McKee, Jean Lapointe, Sofiane Haffouz,
  John~F Weber, David~B Northeast, Philip~J Poole, Geof~C Aers, and Robin~L
  Williams.
\newblock On-chip integration of single photon sources via evanescent coupling
  of tapered nanowires to sin waveguides.
\newblock {\em Advanced Quantum Technologies}, 3(2):1900021, 2020.

\bibitem{chen2016controlling}
Yan Chen, Iman~Esmaeil Zadeh, Klaus~D J{\"o}ns, Andreas Fognini, Michael~E
  Reimer, Jiaxiang Zhang, Dan Dalacu, Philip~J Poole, Fei Ding, Val Zwiller,
  et~al.
\newblock Controlling the exciton energy of a nanowire quantum dot by strain
  fields.
\newblock {\em Applied Physics Letters}, 108(18):182103, 2016.

\bibitem{yang2020proximitized}
Lily Yang, Stephan Steinhauer, Elia Strambini, Thomas Lettner, Lucas
  Schweickert, Marijn~AM Versteegh, Valentina Zannier, Lucia Sorba, Dmitry
  Solenov, and Francesco Giazotto.
\newblock Proximitized josephson junctions in highly-doped inas nanowires
  robust to optical illumination.
\newblock {\em Nanotechnology}, 32(7):075001, 2020.

\bibitem{osada2019strongly}
Alto Osada, Yasutomo Ota, Ryota Katsumi, Masahiro Kakuda, Satoshi Iwamoto, and
  Yasuhiko Arakawa.
\newblock Strongly coupled single-quantum-dot--cavity system integrated on a
  cmos-processed silicon photonic chip.
\newblock {\em Physical Review Applied}, 11(2):024071, 2019.

\bibitem{katsumi2018transfer}
Ryota Katsumi, Yasutomo Ota, Masahiro Kakuda, Satoshi Iwamoto, and Yasuhiko
  Arakawa.
\newblock Transfer-printed single-photon sources coupled to wire waveguides.
\newblock {\em Optica}, 5(6):691--694, 2018.

\bibitem{davanco2017heterogeneous}
Marcelo Davanco, Jin Liu, Luca Sapienza, Chen-Zhao Zhang,
  Jos{\'e}~Vin{\'\i}cius De~Miranda~Cardoso, Varun Verma, Richard Mirin,
  Sae~Woo Nam, Liu Liu, and Kartik Srinivasan.
\newblock Heterogeneous integration for on-chip quantum photonic circuits with
  single quantum dot devices.
\newblock {\em Nature communications}, 8(1):1--12, 2017.

\bibitem{kim2020hybrid}
Je-Hyung Kim, Shahriar Aghaeimeibodi, Jacques Carolan, Dirk Englund, and Edo
  Waks.
\newblock Hybrid integration methods for on-chip quantum photonics.
\newblock {\em Optica}, 7(4):291--308, 2020.

\bibitem{alloing2005growth}
Blandine Alloing, Carl Zinoni, Val Zwiller, LH~Li, Christelle Monat, M~Gobet,
  G~Buchs, Andrea Fiore, E~Pelucchi, and E~Kapon.
\newblock Growth and characterization of single quantum dots emitting at 1300
  nm.
\newblock {\em Applied Physics Letters}, 86(10):101908, 2005.

\bibitem{dalacu2012ultraclean}
Dan Dalacu, Khaled Mnaymneh, Jean Lapointe, Xiaohua Wu, Philip~J Poole,
  Gabriele Bulgarini, Val Zwiller, and Michael~E Reimer.
\newblock Ultraclean emission from inasp quantum dots in defect-free wurtzite
  inp nanowires.
\newblock {\em Nano letters}, 12(11):5919--5923, 2012.

\bibitem{borgstrom2005optically}
Magnus~T Borgstr{\"o}m, Valery Zwiller, Elisabeth M{\"u}ller, and Atac
  Imamoglu.
\newblock Optically bright quantum dots in single nanowires.
\newblock {\em Nano letters}, 5(7):1439--1443, 2005.

\bibitem{deshpande2013electrically}
Saniya Deshpande, Junseok Heo, Ayan Das, and Pallab Bhattacharya.
\newblock Electrically driven polarized single-photon emission from an ingan
  quantum dot in a gan nanowire.
\newblock {\em Nature communications}, 4(1):1--8, 2013.

\bibitem{mrowinski2019excitonic}
Pawe{\l} Mrowi{\'n}ski, Anna Musia{\l}, Krzysztof Gawarecki, {\L}ukasz
  Dusanowski, Tobias Heuser, Nicole Srocka, David Quandt, Andr{\'e}
  Strittmatter, Sven Rodt, Stephan Reitzenstein, et~al.
\newblock Excitonic complexes in mocvd-grown ingaas/gaas quantum dots emitting
  at telecom wavelengths.
\newblock {\em Physical Review B}, 100(11):115310, 2019.

\bibitem{holmes2014room}
Mark~J Holmes, Kihyun Choi, Satoshi Kako, Munetaka Arita, and Yasuhiko Arakawa.
\newblock Room-temperature triggered single photon emission from a iii-nitride
  site-controlled nanowire quantum dot.
\newblock {\em Nano letters}, 14(2):982--986, 2014.

\bibitem{cheng2020epitaxial}
Risheng Cheng, John Wright, Huili~G Xing, Debdeep Jena, and Hong~X Tang.
\newblock Epitaxial niobium nitride superconducting nanowire single-photon
  detectors.
\newblock {\em Applied Physics Letters}, 117(13):132601, 2020.

\bibitem{taylor2021infrared}
Gregor~G Taylor, Dmitry~V Morozov, Ciaran~T Lennon, Peter~S Barry, Calder
  Sheagren, and Robert~H Hadfield.
\newblock Infrared single-photon sensitivity in atomic layer deposited
  superconducting nanowires.
\newblock {\em Applied Physics Letters}, 118(19):191106, 2021.

\bibitem{esmaeil2020efficient}
Iman Esmaeil~Zadeh, Johannes~WN Los, Ronan~BM Gourgues, Jin Chang, Ali~W
  Elshaari, Julien~Romain Zichi, Yuri~J Van~Staaden, Jeroen~PE Swens, Nima
  Kalhor, Antonio Guardiani, et~al.
\newblock Efficient single-photon detection with 7.7 ps time resolution for
  photon-correlation measurements.
\newblock {\em Acs Photonics}, 7(7):1780--1787, 2020.

\bibitem{marsili2013detecting}
Francesco Marsili, Varun~B Verma, Jeffrey~A Stern, Susanmarie Harrington,
  Adriana~E Lita, Thomas Gerrits, Igor Vayshenker, Burm Baek, Matthew~D Shaw,
  Richard~P Mirin, et~al.
\newblock Detecting single infrared photons with 93\% system efficiency.
\newblock {\em Nature Photonics}, 7(3):210--214, 2013.

\bibitem{reddy2020superconducting}
Dileep~V Reddy, Robert~R Nerem, Sae~Woo Nam, Richard~P Mirin, and Varun~B
  Verma.
\newblock Superconducting nanowire single-photon detectors with 98\% system
  detection efficiency at 1550 nm.
\newblock {\em Optica}, 7(12):1649--1653, 2020.

\bibitem{NbRe}
C~Cirillo, J~Chang, M~Caputo, JWN Los, S~Dorenbos, I~Esmaeil~Zadeh, and
  C~Attanasio.
\newblock Superconducting nanowire single photon detectors based on disordered
  nbre films.
\newblock {\em Applied Physics Letters}, 117(17):172602, 2020.

\bibitem{yang2021large}
Can Yang, Mengting Si, Xingyu Zhang, Aobo Yu, Jia Huang, Yiming Pan, Hao Li,
  Lingyun Li, Zhen Wang, Shuo Zhang, et~al.
\newblock Large-area tan superconducting microwire single photon detectors for
  x-ray detection.
\newblock {\em Optics Express}, 29(14):21400--21408, 2021.

\bibitem{hu2020detecting}
Peng Hu, Hao Li, Lixing You, Heqing Wang, You Xiao, Jia Huang, Xiaoyan Yang,
  Weijun Zhang, Zhen Wang, and Xiaoming Xie.
\newblock Detecting single infrared photons toward optimal system detection
  efficiency.
\newblock {\em Optics Express}, 28(24):36884--36891, 2020.

\bibitem{cherednichenko2021low}
Sergey Cherednichenko, Narendra Acharya, Evgenii Novoselov, and Vladimir
  Drakinskiy.
\newblock Low kinetic inductance superconducting mgb2 nanowires with a 130 ps
  relaxation time for single-photon detection applications.
\newblock {\em Superconductor Science and Technology}, 34(4):044001, 2021.

\bibitem{marks1982metalorganic}
J~MARKS, JONL SCHINDLER, and CARL~R KANNEWURF.
\newblock Metalorganic chemical vapor deposition.
\newblock {\em MRS Online Proceedings Library (OPL)}, 1982.

\bibitem{li2016nano}
Jian Li, Robert~A Kirkwood, Luke~J Baker, David Bosworth, Kleanthis
  Erotokritou, Archan Banerjee, Robert~M Heath, Chandra~M Natarajan, Zoe~H
  Barber, Marc Sorel, et~al.
\newblock Nano-optical single-photon response mapping of waveguide integrated
  molybdenum silicide (mosi) superconducting nanowires.
\newblock {\em Optics express}, 24(13):13931--13938, 2016.

\bibitem{oxford2013bs}
Justin~B Spring, Benjamin~J Metcalf, Peter~C Humphreys, W~Steven Kolthammer,
  Xian-Min Jin, Marco Barbieri, Animesh Datta, Nicholas Thomas-Peter, Nathan~K
  Langford, Dmytro Kundys, et~al.
\newblock Boson sampling on a photonic chip.
\newblock {\em Science}, 339(6121):798--801, 2013.

\bibitem{wang2019boson}
Hui Wang, Jian Qin, Xing Ding, Ming-Cheng Chen, Si~Chen, Xiang You, Yu-Ming He,
  Xiao Jiang, L~You, Z~Wang, et~al.
\newblock Boson sampling with 20 input photons and a 60-mode interferometer in
  a 1 0 14-dimensional hilbert space.
\newblock {\em Physical review letters}, 123(25):250503, 2019.

\bibitem{paesani2019generation}
Stefano Paesani, Yunhong Ding, Raffaele Santagati, Levon Chakhmakhchyan,
  Caterina Vigliar, Karsten Rottwitt, Leif~K Oxenl{\o}we, Jianwei Wang, Mark~G
  Thompson, and Anthony Laing.
\newblock Generation and sampling of quantum states of light in a silicon chip.
\newblock {\em Nature Physics}, 15(9):925--929, 2019.

\bibitem{AA2013}
Scott Aaronson and Alex Arkhipov.
\newblock The computational complexity of linear optics.
\newblock {\em Theory of Computing}, 9(4):143--252, 2013.

\bibitem{bsreview}
Daniel~J. Brod, Ernesto~F. Galv{\~a}o, Andrea Crespi, Roberto Osellame,
  Nicol{\`o} Spagnolo, and Fabio Sciarrino.
\newblock {Photonic implementation of boson sampling: a review}.
\newblock {\em Advanced Photonics}, 1(3):034001, 2019.

\bibitem{queensland2013bs}
Matthew~A. Broome, Alessandro Fedrizzi, Saleh Rahimi-Keshari, Justin Dove,
  Scott Aaronson, Timothy~C. Ralph, and Andrew~G. White.
\newblock Photonic boson sampling in a tunable circuit.
\newblock {\em Science}, 339(6121):794--798, 2013.

\bibitem{tillmann2013bs}
Max Tillmann, Borivoje Daki{\'c}, Ren{\'e} Heilmann, Stefan Nolte, Alexander
  Szameit, and Philip Walther.
\newblock Experimental boson sampling.
\newblock {\em Nature photonics}, 7(7):540--544, 2013.

\bibitem{crespi2013bs}
Andrea Crespi, Roberto Osellame, Roberta Ramponi, Daniel~J Brod, Ernesto~F
  Galvao, Nicolo Spagnolo, Chiara Vitelli, Enrico Maiorino, Paolo Mataloni, and
  Fabio Sciarrino.
\newblock Integrated multimode interferometers with arbitrary designs for
  photonic boson sampling.
\newblock {\em Nature photonics}, 7(7):545--549, 2013.

\bibitem{loredo2017boson}
JC~Loredo, MA~Broome, P~Hilaire, O~Gazzano, I~Sagnes, A~Lemaitre, MP~Almeida,
  P~Senellart, and AG~White.
\newblock Boson sampling with single-photon fock states from a bright
  solid-state source.
\newblock {\em Physical review letters}, 118(13):130503, 2017.

\bibitem{hamilton2017gaussian}
Craig~S Hamilton, Regina Kruse, Linda Sansoni, Sonja Barkhofen, Christine
  Silberhorn, and Igor Jex.
\newblock Gaussian boson sampling.
\newblock {\em Physical review letters}, 119(17):170501, 2017.

\bibitem{kruse2019detailed}
Regina Kruse, Craig~S Hamilton, Linda Sansoni, Sonja Barkhofen, Christine
  Silberhorn, and Igor Jex.
\newblock Detailed study of gaussian boson sampling.
\newblock {\em Physical Review A}, 100(3):032326, 2019.

\bibitem{paesani2020near}
Stefano Paesani, Massimo Borghi, Stefano Signorini, Alexandre Ma{\"\i}nos,
  Lorenzo Pavesi, and Anthony Laing.
\newblock Near-ideal spontaneous photon sources in silicon quantum photonics.
\newblock {\em Nature communications}, 11(1):1--6, 2020.

\bibitem{silverstone2015qubit}
Joshua~W Silverstone, Raffaele Santagati, Damien Bonneau, Michael~J Strain,
  Marc Sorel, Jeremy~L O’Brien, and Mark~G Thompson.
\newblock Qubit entanglement between ring-resonator photon-pair sources on a
  silicon chip.
\newblock {\em Nature communications}, 6(1):1--7, 2015.

\bibitem{vaidya2020broadband}
Varun~D Vaidya, B~Morrison, LG~Helt, R~Shahrokshahi, DH~Mahler, MJ~Collins,
  K~Tan, J~Lavoie, A~Repingon, M~Menotti, et~al.
\newblock Broadband quadrature-squeezed vacuum and nonclassical photon number
  correlations from a nanophotonic device.
\newblock {\em Science advances}, 6(39):eaba9186, 2020.

\bibitem{nanophotonics_you}
Lixing You.
\newblock Superconducting nanowire single-photon detectors for quantum
  information.
\newblock {\em Nanophotonics}, 9(9):2673--2692, 2020.

\bibitem{wang2020integrated}
Jianwei Wang, Fabio Sciarrino, Anthony Laing, and Mark~G Thompson.
\newblock Integrated photonic quantum technologies.
\newblock {\em Nature Photonics}, 14(5):273--284, 2020.

\bibitem{huh2015boson}
Joonsuk Huh, Gian~Giacomo Guerreschi, Borja Peropadre, Jarrod~R McClean, and
  Al{\'a}n Aspuru-Guzik.
\newblock Boson sampling for molecular vibronic spectra.
\newblock {\em Nature Photonics}, 9(9):615--620, 2015.

\bibitem{bradler2018gaussian}
Kamil Br{\'a}dler, Pierre-Luc Dallaire-Demers, Patrick Rebentrost, Daiqin Su,
  and Christian Weedbrook.
\newblock Gaussian boson sampling for perfect matchings of arbitrary graphs.
\newblock {\em Physical Review A}, 98(3):032310, 2018.

\bibitem{arrazola2018using}
Juan~Miguel Arrazola and Thomas~R Bromley.
\newblock Using gaussian boson sampling to find dense subgraphs.
\newblock {\em Physical review letters}, 121(3):030503, 2018.

\bibitem{banchi2020molecular}
Leonardo Banchi, Mark Fingerhuth, Tomas Babej, Christopher Ing, and Juan~Miguel
  Arrazola.
\newblock Molecular docking with gaussian boson sampling.
\newblock {\em Science advances}, 6(23):eaax1950, 2020.

\bibitem{arrazola2021quantum}
Juan~M Arrazola, Ville Bergholm, Kamil Br{\'a}dler, Thomas~R Bromley, Matt~J
  Collins, Ish Dhand, Alberto Fumagalli, Thomas Gerrits, Andrey Goussev,
  Lukas~G Helt, et~al.
\newblock Quantum circuits with many photons on a programmable nanophotonic
  chip.
\newblock {\em Nature}, 591(7848):54--60, 2021.

\bibitem{modular-interferometer}
Paolo~L Mennea, William~R Clements, Devin~H Smith, James~C Gates, Benjamin~J
  Metcalf, Rex~HS Bannerman, Roel Burgwal, Jelmer~J Renema, W~Steven
  Kolthammer, Ian~A Walmsley, et~al.
\newblock Modular linear optical circuits.
\newblock {\em Optica}, 5(9):1087--1090, 2018.

\bibitem{tang2018fs}
Hao Tang, Carlo Di~Franco, Zi-Yu Shi, Tian-Shen He, Zhen Feng, Jun Gao, Ke~Sun,
  Zhan-Ming Li, Zhi-Qiang Jiao, Tian-Yu Wang, et~al.
\newblock Experimental quantum fast hitting on hexagonal graphs.
\newblock {\em Nature Photonics}, 12(12):754--758, 2018.

\bibitem{peruzzo2010quantum}
Alberto Peruzzo, Mirko Lobino, Jonathan~CF Matthews, Nobuyuki Matsuda, Alberto
  Politi, Konstantinos Poulios, Xiao-Qi Zhou, Yoav Lahini, Nur Ismail, Kerstin
  W{\"o}rhoff, et~al.
\newblock Quantum walks of correlated photons.
\newblock {\em Science}, 329(5998):1500--1503, 2010.

\bibitem{quantum_walksQiang}
Xiaogang Qiang, Yizhi Wang, Shichuan Xue, Renyou Ge, Lifeng Chen, Yingwen Liu,
  Anqi Huang, Xiang Fu, Ping Xu, Teng Yi, et~al.
\newblock Implementing graph-theoretic quantum algorithms on a silicon photonic
  quantum walk processor.
\newblock {\em Science Advances}, 7(9):eabb8375, 2021.

\bibitem{quantum_walksWang}
Yizhi Wang, Yingwen Liu, Junwei Zhan, Shichuan Xue, Yuzhen Zheng, Ru~Zeng,
  Zhihao Wu, Zihao Wang, Qilin Zheng, Dongyang Wang, et~al.
\newblock Large-scale full-programmable quantum walk and its applications.
\newblock {\em arXiv preprint arXiv:2208.13186}, 2022.

\bibitem{aharonov1993quantum}
Yakir Aharonov, Luiz Davidovich, and Nicim Zagury.
\newblock Quantum random walks.
\newblock {\em Physical Review A}, 48(2):1687, 1993.

\bibitem{childs2003exponential}
Andrew~M Childs, Richard Cleve, Enrico Deotto, Edward Farhi, Sam Gutmann, and
  Daniel~A Spielman.
\newblock Exponential algorithmic speedup by a quantum walk.
\newblock In {\em Proceedings of the thirty-fifth annual ACM symposium on
  Theory of computing}, pages 59--68, 2003.

\bibitem{childs2004spatial}
Andrew~M Childs and Jeffrey Goldstone.
\newblock Spatial search by quantum walk.
\newblock {\em Physical Review A}, 70(2):022314, 2004.

\bibitem{benedetti2021quantum}
Claudia Benedetti, Dario Tamascelli, Matteo~GA Paris, and Andrea Crespi.
\newblock Quantum spatial search in two-dimensional waveguide arrays.
\newblock {\em Physical Review Applied}, 16(5):054036, 2021.

\bibitem{quantum_walks_review}
Julia Kempe.
\newblock Quantum random walks: an introductory overview.
\newblock {\em Contemporary Physics}, 44(4):307--327, 2003.

\bibitem{childs2009universal}
Andrew~M Childs.
\newblock Universal computation by quantum walk.
\newblock {\em Physical review letters}, 102(18):180501, 2009.

\bibitem{childs2013universal}
Andrew~M Childs, David Gosset, and Zak Webb.
\newblock Universal computation by multiparticle quantum walk.
\newblock {\em Science}, 339(6121):791--794, 2013.

\bibitem{broome2010discrete}
Matthew~A Broome, Alessandro Fedrizzi, Benjimain~P Lanyon, Ivan Kassal, Alan
  Aspuru-Guzik, and Andrew~G White.
\newblock Discrete single-photon quantum walks with tunable decoherence.
\newblock {\em Physical review letters}, 104(15):153602, 2010.

\bibitem{schreiber2010photons}
Andreas Schreiber, Katiuscia~N Cassemiro, V{\'a}clav Poto{\v{c}}ek, Aur{\'e}l
  G{\'a}bris, Peter~James Mosley, Erika Andersson, Igor Jex, and Ch~Silberhorn.
\newblock Photons walking the line: a quantum walk with adjustable coin
  operations.
\newblock {\em Physical review letters}, 104(5):050502, 2010.

\bibitem{schreiber20122d}
Andreas Schreiber, Aur{\'e}l G{\'a}bris, Peter~P Rohde, Kaisa Laiho, Martin
  {\v{S}}tefa{\v{n}}{\'a}k, V{\'a}clav Poto{\v{c}}ek, Craig Hamilton, Igor Jex,
  and Christine Silberhorn.
\newblock A 2d quantum walk simulation of two-particle dynamics.
\newblock {\em Science}, 336(6077):55--58, 2012.

\bibitem{tang2018experimental}
Hao Tang, Xiao-Feng Lin, Zhen Feng, Jing-Yuan Chen, Jun Gao, Ke~Sun, Chao-Yue
  Wang, Peng-Cheng Lai, Xiao-Yun Xu, Yao Wang, et~al.
\newblock Experimental two-dimensional quantum walk on a photonic chip.
\newblock {\em Science advances}, 4(5):eaat3174, 2018.

\bibitem{jiao2021two}
Zhi-Qiang Jiao, Jun Gao, Wen-Hao Zhou, Xiao-Wei Wang, Ruo-Jing Ren, Xiao-Yun
  Xu, Lu-Feng Qiao, Yao Wang, and Xian-Min Jin.
\newblock Two-dimensional quantum walks of correlated photons.
\newblock {\em Optica}, 8(9):1129--1135, 2021.

\bibitem{sibson2017chip}
Philip Sibson, Chris Erven, Mark Godfrey, Shigehito Miki, Taro Yamashita, Mikio
  Fujiwara, Masahide Sasaki, Hirotaka Terai, Michael~G Tanner, Chandra~M
  Natarajan, et~al.
\newblock Chip-based quantum key distribution.
\newblock {\em Nature communications}, 8(1):1--6, 2017.

\bibitem{semenenko2020chip}
Henry Semenenko, Philip Sibson, Andy Hart, Mark~G Thompson, John~G Rarity, and
  Chris Erven.
\newblock Chip-based measurement-device-independent quantum key distribution.
\newblock {\em Optica}, 7(3):238--242, 2020.

\bibitem{beutel2021detector}
Fabian Beutel, Helge Gehring, Martin~A Wolff, Carsten Schuck, and Wolfram
  Pernice.
\newblock Detector-integrated on-chip qkd receiver for ghz clock rates.
\newblock {\em npj Quantum Information}, 7(1):1--8, 2021.

\bibitem{QKD_optica}
Fabian Beutel, Frank Br{\"u}ckerhoff-Pl{\"u}ckelmann, Helge Gehring, Vadim
  Kovalyuk, Philipp Zolotov, Gregory Goltsman, and Wolfram~HP Pernice.
\newblock Fully integrated four-channel wavelength-division multiplexed qkd
  receiver.
\newblock {\em Optica}, 9(10):1121--1130, 2022.

\bibitem{yin2017satellite}
Juan Yin, Yuan Cao, Yu-Huai Li, Sheng-Kai Liao, Liang Zhang, Ji-Gang Ren,
  Wen-Qi Cai, Wei-Yue Liu, Bo~Li, Hui Dai, et~al.
\newblock Satellite-based entanglement distribution over 1200 kilometers.
\newblock {\em Science}, 356(6343):1140--1144, 2017.

\bibitem{liao2017satellite}
Sheng-Kai Liao, Wen-Qi Cai, Wei-Yue Liu, Liang Zhang, Yang Li, Ji-Gang Ren,
  Juan Yin, Qi~Shen, Yuan Cao, Zheng-Ping Li, et~al.
\newblock Satellite-to-ground quantum key distribution.
\newblock {\em Nature}, 549(7670):43--47, 2017.

\bibitem{ren2017ground}
Ji-Gang Ren, Ping Xu, Hai-Lin Yong, Liang Zhang, Sheng-Kai Liao, Juan Yin,
  Wei-Yue Liu, Wen-Qi Cai, Meng Yang, Li~Li, et~al.
\newblock Ground-to-satellite quantum teleportation.
\newblock {\em Nature}, 549(7670):70--73, 2017.

\bibitem{liao2018satellite}
Sheng-Kai Liao, Wen-Qi Cai, Johannes Handsteiner, Bo~Liu, Juan Yin, Liang
  Zhang, Dominik Rauch, Matthias Fink, Ji-Gang Ren, Wei-Yue Liu, et~al.
\newblock Satellite-relayed intercontinental quantum network.
\newblock {\em Physical review letters}, 120(3):030501, 2018.

\bibitem{paraiso2021photonic}
Taofiq~K Paraiso, Thomas Roger, Davide~G Marangon, Innocenzo De~Marco, Mirko
  Sanzaro, Robert~I Woodward, James~F Dynes, Zhiliang Yuan, and Andrew~J
  Shields.
\newblock A photonic integrated quantum secure communication system.
\newblock {\em Nature Photonics}, 15(11):850--856, 2021.

\bibitem{ma2016silicon}
Chaoxuan Ma, Wesley~D Sacher, Zhiyuan Tang, Jared~C Mikkelsen, Yisu Yang, Feihu
  Xu, Torrey Thiessen, Hoi-Kwong Lo, and Joyce~KS Poon.
\newblock Silicon photonic transmitter for polarization-encoded quantum key
  distribution.
\newblock {\em Optica}, 3(11):1274--1278, 2016.

\bibitem{sibson2017integrated}
Philip Sibson, Jake~E Kennard, Stasja Stanisic, Chris Erven, Jeremy~L
  O’Brien, and Mark~G Thompson.
\newblock Integrated silicon photonics for high-speed quantum key distribution.
\newblock {\em Optica}, 4(2):172--177, 2017.

\bibitem{ding2017high}
Yunhong Ding, Davide Bacco, Kjeld Dalgaard, Xinlun Cai, Xiaoqi Zhou, Karsten
  Rottwitt, and Leif~Katsuo Oxenl{\o}we.
\newblock High-dimensional quantum key distribution based on multicore fiber
  using silicon photonic integrated circuits.
\newblock {\em npj Quantum Information}, 3(1):1--7, 2017.

\bibitem{bunandar2018metropolitan}
Darius Bunandar, Anthony Lentine, Catherine Lee, Hong Cai, Christopher~M Long,
  Nicholas Boynton, Nicholas Martinez, Christopher DeRose, Changchen Chen,
  Matthew Grein, et~al.
\newblock Metropolitan quantum key distribution with silicon photonics.
\newblock {\em Physical Review X}, 8(2):021009, 2018.

\bibitem{zhang2019integrated}
Gong Zhang, Jing~Yan Haw, Hong Cai, Feng Xu, SM~Assad, Joseph~F Fitzsimons,
  Xianzhong Zhou, Y~Zhang, S~Yu, J~Wu, et~al.
\newblock An integrated silicon photonic chip platform for continuous-variable
  quantum key distribution.
\newblock {\em Nature Photonics}, 13(12):839--842, 2019.

\bibitem{llewellyn2020chip}
Daniel Llewellyn, Yunhong Ding, Imad~I Faruque, Stefano Paesani, Davide Bacco,
  Raffaele Santagati, Yan-Jun Qian, Yan Li, Yun-Feng Xiao, Marcus Huber, et~al.
\newblock Chip-to-chip quantum teleportation and multi-photon entanglement in
  silicon.
\newblock {\em Nature Physics}, 16(2):148--153, 2020.

\bibitem{QKD_RMP1}
Feihu Xu, Xiongfeng Ma, Qiang Zhang, Hoi-Kwong Lo, and Jian-Wei Pan.
\newblock Secure quantum key distribution with realistic devices.
\newblock {\em Reviews of Modern Physics}, 92(2):025002, 2020.

\bibitem{QKD_RMP2}
Chao-Yang Lu, Yuan Cao, Cheng-Zhi Peng, and Jian-Wei Pan.
\newblock Micius quantum experiments in space.
\newblock {\em Reviews of Modern Physics}, 94(3):035001, 2022.

\bibitem{silver2017mastering}
David Silver, Julian Schrittwieser, Karen Simonyan, Ioannis Antonoglou, Aja
  Huang, Arthur Guez, Thomas Hubert, Lucas Baker, Matthew Lai, Adrian Bolton,
  et~al.
\newblock Mastering the game of go without human knowledge.
\newblock {\em nature}, 550(7676):354--359, 2017.

\bibitem{sengupta2014power}
Biswa Sengupta and Martin~B Stemmler.
\newblock Power consumption during neuronal computation.
\newblock {\em Proceedings of the IEEE}, 102(5):738--750, 2014.

\bibitem{schwabe1998electronic}
RJ~Schwabe, S~Zelinger, TS~Key, and KO~Phipps.
\newblock Electronic lighting interference.
\newblock {\em IEEE industry applications magazine}, 4(4):43--48, 1998.

\bibitem{wetzstein2020inference}
Gordon Wetzstein, Aydogan Ozcan, Sylvain Gigan, Shanhui Fan, Dirk Englund,
  Marin Solja{\v{c}}i{\'c}, Cornelia Denz, David~AB Miller, and Demetri
  Psaltis.
\newblock Inference in artificial intelligence with deep optics and photonics.
\newblock {\em Nature}, 588(7836):39--47, 2020.

\bibitem{liu2021research}
Jia Liu, Qiuhao Wu, Xiubao Sui, Qian Chen, Guohua Gu, Liping Wang, and Shengcai
  Li.
\newblock Research progress in optical neural networks: theory, applications
  and developments.
\newblock {\em PhotoniX}, 2(1):1--39, 2021.

\bibitem{zhang2021optical}
H~Zhang, M~Gu, XD~Jiang, J~Thompson, H~Cai, S~Paesani, R~Santagati, A~Laing,
  Y~Zhang, MH~Yung, et~al.
\newblock An optical neural chip for implementing complex-valued neural
  network.
\newblock {\em Nature communications}, 12(1):1--11, 2021.

\bibitem{shen2017deep}
Yichen Shen, Nicholas~C Harris, Scott Skirlo, Mihika Prabhu, Tom Baehr-Jones,
  Michael Hochberg, Xin Sun, Shijie Zhao, Hugo Larochelle, Dirk Englund, et~al.
\newblock Deep learning with coherent nanophotonic circuits.
\newblock {\em Nature photonics}, 11(7):441--446, 2017.

\bibitem{shainline2017superconducting}
Jeffrey~M Shainline, Sonia~M Buckley, Richard~P Mirin, and Sae~Woo Nam.
\newblock Superconducting optoelectronic circuits for neuromorphic computing.
\newblock {\em Physical Review Applied}, 7(3):034013, 2017.

\bibitem{zhu2022space}
HH~Zhu, J~Zou, H~Zhang, YZ~Shi, SB~Luo, N~Wang, H~Cai, LX~Wan, B~Wang,
  XD~Jiang, et~al.
\newblock Space-efficient optical computing with an integrated chip diffractive
  neural network.
\newblock {\em Nature communications}, 13(1):1--9, 2022.

\bibitem{lau2022photonic}
Jonathan Wei~Zhong Lau, Hui Zhang, Lingxiao Wan, Liang Shi, Hong Cai, Xianshu
  Luo, Patrick Lo, Chee-Kong Lee, Leong-Chuan Kwek, and Ai~Qun Liu.
\newblock A photonic chip-based machine learning approach for the prediction of
  molecular properties.
\newblock {\em arXiv preprint arXiv:2203.02285}, 2022.

\bibitem{yan2022all}
Tao Yan, Rui Yang, Ziyang Zheng, Xing Lin, Hongkai Xiong, and Qionghai Dai.
\newblock All-optical graph representation learning using integrated
  diffractive photonic computing units.
\newblock {\em Science Advances}, 8(24):eabn7630, 2022.

\bibitem{mourgias2022noise}
G~Mourgias-Alexandris, M~Moralis-Pegios, A~Tsakyridis, S~Simos, G~Dabos,
  A~Totovic, N~Passalis, M~Kirtas, T~Rutirawut, FY~Gardes, et~al.
\newblock Noise-resilient and high-speed deep learning with coherent silicon
  photonics.
\newblock {\em Nature communications}, 13(1):1--7, 2022.

\bibitem{cong2022chip}
Guangwei Cong, Noritsugu Yamamoto, Takashi Inoue, Yuriko Maegami, Morifumi
  Ohno, Shota Kita, Shu Namiki, and Koji Yamada.
\newblock On-chip bacterial foraging training in silicon photonic circuits for
  projection-enabled nonlinear classification.
\newblock {\em Nature communications}, 13(1):1--12, 2022.

\bibitem{ashtiani2022chip}
Farshid Ashtiani, Alexander~J Geers, and Firooz Aflatouni.
\newblock An on-chip photonic deep neural network for image classification.
\newblock {\em Nature}, pages 1--6, 2022.

\bibitem{bandyopadhyay2022single}
Saumil Bandyopadhyay, Alexander Sludds, Stefan Krastanov, Ryan Hamerly,
  Nicholas Harris, Darius Bunandar, Matthew Streshinsky, Michael Hochberg, and
  Dirk Englund.
\newblock Single chip photonic deep neural network with accelerated training.
\newblock {\em arXiv preprint arXiv:2208.01623}, 2022.

\bibitem{buckley2020integrated}
Sonia~M Buckley, Alexander~N Tait, Jeffrey Chiles, Adam~N McCaughan, Saeed
  Khan, Richard~P Mirin, Sae~Woo Nam, and Jeffrey~M Shainline.
\newblock Integrated-photonic characterization of single-photon detectors for
  use in neuromorphic synapses.
\newblock {\em Physical Review Applied}, 14(5):054008, 2020.

\bibitem{schuld2015introduction}
Maria Schuld, Ilya Sinayskiy, and Francesco Petruccione.
\newblock An introduction to quantum machine learning.
\newblock {\em Contemporary Physics}, 56(2):172--185, 2015.

\bibitem{biamonte2017quantum}
Jacob Biamonte, Peter Wittek, Nicola Pancotti, Patrick Rebentrost, Nathan
  Wiebe, and Seth Lloyd.
\newblock Quantum machine learning.
\newblock {\em Nature}, 549(7671):195--202, 2017.

\bibitem{haug2021large}
Tobias Haug, Chris~N Self, and MS~Kim.
\newblock Large-scale quantum machine learning.
\newblock {\em arXiv preprint arXiv:2108.01039}, 2021.

\bibitem{huang2021power}
Hsin-Yuan Huang, Michael Broughton, Masoud Mohseni, Ryan Babbush, Sergio Boixo,
  Hartmut Neven, and Jarrod~R McClean.
\newblock Power of data in quantum machine learning.
\newblock {\em Nature communications}, 12(1):1--9, 2021.

\bibitem{krenn2020computer}
Mario Krenn, Manuel Erhard, and Anton Zeilinger.
\newblock Computer-inspired quantum experiments.
\newblock {\em Nature Reviews Physics}, 2(11):649--661, 2020.

\bibitem{lamata2020quantum}
Lucas Lamata.
\newblock Quantum machine learning and quantum biomimetics: A perspective.
\newblock {\em Machine Learning: Science and Technology}, 1(3):033002, 2020.

\bibitem{cerezo2022challenges}
M~Cerezo, Guillaume Verdon, Hsin-Yuan Huang, Lukasz Cincio, and Patrick~J
  Coles.
\newblock Challenges and opportunities in quantum machine learning.
\newblock {\em Nature Computational Science}, 2(9):567--576, 2022.

\bibitem{lidar1550}
Haiyun Xia, Guoliang Shentu, Mingjia Shangguan, Xiuxiu Xia, Xiaodong Jia, Chong
  Wang, Jun Zhang, Jason~S Pelc, MM~Fejer, Qiang Zhang, et~al.
\newblock Long-range micro-pulse aerosol lidar at 1.5 $\mu$m with an
  upconversion single-photon detector.
\newblock {\em Optics letters}, 40(7):1579--1582, 2015.

\bibitem{lidar2}
Anu Swatantran, Hao Tang, Terence Barrett, Phil DeCola, and Ralph Dubayah.
\newblock Rapid, high-resolution forest structure and terrain mapping over
  large areas using single photon lidar.
\newblock {\em Scientific reports}, 6(1):1--12, 2016.

\bibitem{lidar3}
Jiang Zhu, Yajun Chen, Labao Zhang, Xiaoqing Jia, Zhijun Feng, Ganhua Wu,
  Xiachao Yan, Jiquan Zhai, Yang Wu, Qi~Chen, et~al.
\newblock Demonstration of measuring sea fog with an snspd-based lidar system.
\newblock {\em Scientific reports}, 7(1):1--7, 2017.

\bibitem{lidar4}
Zheng-Ping Li, Jun-Tian Ye, Xin Huang, Peng-Yu Jiang, Yuan Cao, Yu~Hong, Chao
  Yu, Jun Zhang, Qiang Zhang, Cheng-Zhi Peng, et~al.
\newblock Single-photon imaging over 200 km.
\newblock {\em Optica}, 8(3):344--349, 2021.

\bibitem{lidar-review1}
Yanqiu Guan, Haochen Li, Li~Xue, Rui Yin, Labao Zhang, Hao Wang, Guanghao Zhu,
  Lin Kang, Jian Chen, and Peiheng Wu.
\newblock Lidar with superconducting nanowire single-photon detectors: Recent
  advances and developments.
\newblock {\em Optics and Lasers in Engineering}, 156:107102, 2022.

\bibitem{mid-infrared}
Jin Chang, Johannes~WN Los, Ronan Gourgues, Stephan Steinhauer, SN~Dorenbos,
  Silvania~F Pereira, H~Paul Urbach, Val Zwiller, and Iman~Esmaeil Zadeh.
\newblock Efficient mid-infrared single-photon detection using superconducting
  nbtin nanowires with high time resolution in a gifford-mcmahon cryocooler.
\newblock {\em Photonics Research}, 10(4):1063--1070, 2022.

\bibitem{lidar5}
Aude Martin, Delphin Dodane, Luc Leviandier, Daniel Dolfi, Alan Naughton, Peter
  O’Brien, Thijs Spuessens, Roel Baets, Guy Lepage, Peter Verheyen, et~al.
\newblock Photonic integrated circuit-based fmcw coherent lidar.
\newblock {\em Journal of Lightwave Technology}, 36(19):4640--4645, 2018.

\bibitem{lidar6}
David~M Winker, Richard~H Couch, and MPatrick McCormick.
\newblock An overview of lite: Nasa's lidar in-space technology experiment.
\newblock {\em Proceedings of the IEEE}, 84(2):164--180, 1996.

\bibitem{meta1}
Shengyuan Chang, Xuexue Guo, and Xingjie Ni.
\newblock Optical metasurfaces: progress and applications.
\newblock {\em Annu. Rev. Mater. Res}, 48(1):279--302, 2018.

\bibitem{meta2}
Hou-Tong Chen, Antoinette~J Taylor, and Nanfang Yu.
\newblock A review of metasurfaces: physics and applications.
\newblock {\em Reports on progress in physics}, 79(7):076401, 2016.

\bibitem{meta3}
Romain Fleury, Dimitrios~L Sounas, and Andrea Alu.
\newblock Negative refraction and planar focusing based on parity-time
  symmetric metasurfaces.
\newblock {\em Physical review letters}, 113(2):023903, 2014.

\bibitem{meta4}
Mohammadreza Khorasaninejad, Francesco Aieta, Pritpal Kanhaiya, Mikhail~A Kats,
  Patrice Genevet, David Rousso, and Federico Capasso.
\newblock Achromatic metasurface lens at telecommunication wavelengths.
\newblock {\em Nano letters}, 15(8):5358--5362, 2015.

\bibitem{meta5}
Yihao Yang, Huaping Wang, Faxin Yu, Zhiwei Xu, and Hongsheng Chen.
\newblock A metasurface carpet cloak for electromagnetic, acoustic and water
  waves.
\newblock {\em Scientific reports}, 6(1):1--6, 2016.

\bibitem{meta-QDs}
Yanjun Bao, Qiaoling Lin, Rongbin Su, Zhang-Kai Zhou, Jindong Song, Juntao Li,
  and Xue-Hua Wang.
\newblock On-demand spin-state manipulation of single-photon emission from
  quantum dot integrated with metasurface.
\newblock {\em Science advances}, 6(31):eaba8761, 2020.

\bibitem{vaskin2019light}
Aleksandr Vaskin, Radoslaw Kolkowski, A~Femius Koenderink, and Isabelle Staude.
\newblock Light-emitting metasurfaces.
\newblock {\em Nanophotonics}, 8(7):1151--1198, 2019.

\bibitem{zheng2022photon}
Jingyuan Zheng, You Xiao, Mingzhong Hu, Yuchen Zhao, Hao Li, Lixing You, Xue
  Feng, Fang Liu, Kaiyu Cui, Yidong Huang, et~al.
\newblock A photon counting reconstructive spectrometer combining metasurfaces
  and superconducting nanowire single-photon detectors.
\newblock {\em arXiv preprint arXiv:2207.09234}, 2022.

\bibitem{xiao2022superconducting}
You Xiao, Shuai Wei, Jiajia Xu, Ruoyan Ma, Xiaoyu Liu, Xiaofu Zhang, Tiger~H
  Tao, Hao Li, Zengqi Wang, Lixing You, et~al.
\newblock Superconducting single-photon spectrometer with 3d-printed
  photonic-crystal filters.
\newblock {\em ACS Photonics}, 2022.

\bibitem{meta-review1}
Alexander~S Solntsev, Girish~S Agarwal, and Yuri~S Kivshar.
\newblock Metasurfaces for quantum photonics.
\newblock {\em Nature Photonics}, 15(5):327--336, 2021.

\bibitem{pagani2014tunable}
Mattia Pagani, David Marpaung, Duk-Yong Choi, Steve~J Madden, Barry
  Luther-Davies, and Benjamin~J Eggleton.
\newblock Tunable wideband microwave photonic phase shifter using on-chip
  stimulated brillouin scattering.
\newblock {\em Optics express}, 22(23):28810--28818, 2014.

\bibitem{russell2012cryogenic}
Damon Russell, Kieran Cleary, and Rodrigo Reeves.
\newblock Cryogenic probe station for on-wafer characterization of electrical
  devices.
\newblock {\em Review of Scientific Instruments}, 83(4):044703, 2012.

\bibitem{pan2021transfer}
Jieming Pan, Kain~Lu Low, Joydeep Ghosh, Senthilnath Jayavelu, Md~Meftahul
  Ferdaus, Shang~Yi Lim, Evgeny Zamburg, Yida Li, Baoshan Tang, Xinghua Wang,
  et~al.
\newblock Transfer learning-based artificial intelligence-integrated physical
  modeling to enable failure analysis for 3 nanometer and smaller silicon-based
  cmos transistors.
\newblock {\em ACS Applied Nano Materials}, 4(7):6903--6915, 2021.

\bibitem{Michalet_2005_QD_Review}
X.~Michalet, F.~F. Pinaud, L.~A. Bentolila, J.~M. Tsay, S.~Doose, J.~J. Li,
  G.~Sundaresan, A.~M. Wu, S.~S. Gambhir, and S.~Weiss.
\newblock Quantum dots for live cells, in vivo imaging, and diagnostics.
\newblock {\em Science}, 307(5709):538--544, 2005.

\bibitem{Chinnathambi:2019_QD_bio_review}
Shanmugavel Chinnathambi and Naoto Shirahata.
\newblock Recent advances on fluorescent biomarkers of near-infrared quantum
  dots for in vitro and in vivo imaging.
\newblock {\em Science and Technology of Advanced Materials}, 20(1):337--355,
  2019.
\newblock PMID: 31068983.

\bibitem{Jingjing_2013_QD_bio_review}
Jingjing Li and Jun-Jie Zhu.
\newblock Quantum dots for fluorescent biosensing and bio-imaging applications.
\newblock {\em Analyst}, 138:2506--2515, 2013.

\bibitem{FARZIN2021_QD_bio_review}
Mohammad~Ali Farzin and Hassan Abdoos.
\newblock A critical review on quantum dots: From synthesis toward applications
  in electrochemical biosensors for determination of disease-related
  biomolecules.
\newblock {\em Talanta}, 224:121828, 2021.

\bibitem{Boissiere2013_QD_bio_review}
Michel Boissiere.
\newblock {\em Quantum Dots as Biomarker}, pages 75--97.
\newblock Springer London, London, 2013.

\bibitem{Brunetti2018_infrared_QD_imaging}
Jlenia Brunetti, Giulia Riolo, Mariangela Gentile, Andrea Bernini, Eugenio
  Paccagnini, Chiara Falciani, Luisa Lozzi, Silvia Scali, Lorenzo Depau,
  Alessandro Pini, Pietro Lupetti, and Luisa Bracci.
\newblock Near-infrared quantum dots labelled with a tumor selective
  tetrabranched peptide for in vivo imaging.
\newblock {\em Journal of Nanobiotechnology}, 16(1):21, Mar 2018.

\bibitem{GIL2021_infrared_QD_for_bio}
Hélio~M. Gil, Thomas~W. Price, Kanik Chelani, Jean-Sebastien~G. Bouillard,
  Simon~D.J. Calaminus, and Graeme~J. Stasiuk.
\newblock Nir-quantum dots in biomedical imaging and their future.
\newblock {\em iScience}, 24(3):102189, 2021.

\bibitem{Fei_QD_SNSPD_brainimaging_1300_2021}
Fei Xia, Monique Gevers, Andreas Fognini, Aaron~T. Mok, Bo~Li, Najva Akbari,
  Iman~Esmaeil Zadeh, Jessie Qin-Dregely, and Chris Xu.
\newblock Short-wave infrared confocal fluorescence imaging of deep mouse brain
  with a superconducting nanowire single-photon detector.
\newblock {\em ACS Photonics}, 8(9):2800--2810, 2021.

\bibitem{Wang_brain_QD_SNSPD_3rdopticalwindow_2022}
Feifei Wang, Fuqiang Ren, Zhuoran Ma, Liangqiong Qu, Ronan Gourgues, Chun Xu,
  Ani Baghdasaryan, Jiachen Li, Iman~Esmaeil Zadeh, Johannes W.~N. Los, Andreas
  Fognini, Jessie Qin-Dregely, and Hongjie Dai.
\newblock In vivo non-invasive confocal fluorescence imaging beyond
  1,700{\thinspace}nm using superconducting nanowire single-photon detectors.
\newblock {\em Nature Nanotechnology}, 17(6):653--660, Jun 2022.

\bibitem{ozcelik2015optofluidic}
Damla Ozcelik, Joshua~W Parks, Thomas~A Wall, Matthew~A Stott, Hong Cai,
  Joseph~W Parks, Aaron~R Hawkins, and Holger Schmidt.
\newblock Optofluidic wavelength division multiplexing for single-virus
  detection.
\newblock {\em Proceedings of the National Academy of Sciences},
  112(42):12933--12937, 2015.

\bibitem{pelton2000spectroscopic}
John~T Pelton and Larry~R McLean.
\newblock Spectroscopic methods for analysis of protein secondary structure.
\newblock {\em Analytical biochemistry}, 277(2):167--176, 2000.

\bibitem{pieczonka2008single}
Nicholas~PW Pieczonka and Ricardo~F Aroca.
\newblock Single molecule analysis by surfaced-enhanced raman scattering.
\newblock {\em chemical society reviews}, 37(5):946--954, 2008.

\end{thebibliography}
\end{document}